\documentclass[%
secnumarabic,
 preprint,
 notitlepage,
superscriptaddress,
 amsmath,amssymb,
 aps,
sns-aps,
onecolumn,
nofootinbib,
nobibnotes
]{revtex4-2}

\usepackage{float}

\usepackage{hyperref}
\usepackage{hhline}
\usepackage{cancel}
\usepackage{tikz} 
\usepackage{mathrsfs}

\usetikzlibrary{plotmarks}

\usepackage{xcolor}

\usepackage{graphicx,caption,subcaption}
\usetikzlibrary{decorations.pathmorphing}

\usepackage{dcolumn}
\usepackage{bm}
\usepackage[mathlines]{lineno}

\usepackage{url}
\usepackage{savesym}
\usepackage{mathrsfs}
\savesymbol{pdfbookmark}
\usepackage{hyperref}

\usepackage{tikz} 
\usepackage{tikz-feynman}
\tikzfeynmanset{compat=1.0.0}
\usetikzlibrary{shapes,arrows,positioning,automata,backgrounds,calc,er,patterns}

\usepackage{simplewick}

\usepackage[utf8]{inputenc} 
\usepackage[russian,english]{babel}

\usepackage{graphicx}
\usepackage{epstopdf}

\usepackage{hyperref}
\usepackage{fancyhdr}
\usepackage{lastpage}
\usepackage[titletoc]{appendix}
\usepackage[OT2, T1]{fontenc}
\usepackage[scaled]{beramono}
\usepackage{listings}
\usepackage{upquote}
\usepackage{setspace}
\usepackage{bm}
\definecolor{gray}{gray}{0.5}
\usepackage{empheq}

\usepackage[export]{adjustbox}

\usepackage{mathtools}

\usepackage{hyperref}

\usepackage{mwe}

\usepackage{caption}

\renewcommand{\thesection}{}

\makeatletter
\def\@seccntformat#1{\csname #1ignore\expandafter\endcsname\csname the#1\endcsname\quad}
\let\sectionignore\@gobbletwo
\let\latex@numberline\numberline
\def\numberline#1{\if\relax#1\relax\else\latex@numberline{#1}\fi}
\makeatother
\usepackage{slashbox}
\usepackage{titlesec}
\usepackage{blindtext}
\usepackage [english]{babel}
\usepackage [autostyle, english = american]{csquotes}
\MakeOuterQuote{"}

\allowdisplaybreaks

\usepackage{titlesec}
\titleformat{\section}
  {\normalfont\sffamily\large\bfseries\color{black}}
  {\thesection}{1em}{}

\hypersetup{colorlinks=true,urlcolor=blue}

\usepackage{tabularx}

\usepackage[final]{changes}

\usepackage{lipsum}
\definechangesauthor[color=blue]{.}
\colorlet{Changes@Color}{red}

\begin{document}


\title{Signatures of a Majorana-Fermi surface in the Kitaev magnet Ag$_3$LiIr$_2$O$_6$}



\author{Joshuah T. Heath}

\affiliation{Department of Physics and Astronomy, Dartmouth College, 6127 Wilder Laboratory, Hanover, New Hampshire 03755, USA}
\email{Joshuah.T.Heath@dartmouth.edu}

\affiliation{%
Physics Department,  Boston  College,  Chestnut  Hill, Massachusetts  02467,  USA
}%


\author{Faranak Bahrami}

\affiliation{%
Physics Department,  Boston  College,  Chestnut  Hill, Massachusetts  02467,  USA
}%

\author{Sangyun Lee}
\affiliation{ 
MPA-Q,  Los  Alamos  National  Laboratory,  Los  Alamos,  New Mexico  87545, USA
}

\author{Roman Movshovich}
\affiliation{ 
MPA-Q,  Los  Alamos  National  Laboratory,  Los  Alamos,  New Mexico  87545, USA
}

\author{Xiao Chen}

 \author{Fazel Tafti}

\author{Kevin S. Bedell}%

\affiliation{%
Physics Department,  Boston  College,  Chestnut  Hill, Massachusetts  02467,  USA
}%


%




\date{\today}


\begin{abstract}
Detecting Majorana fermions in experimental realizations of the Kitaev honeycomb model is often complicated by non-trivial interactions inherent to potential spin liquid candidates.
In this {work}, we {identify} several distinct thermodynamic signatures of massive, itinerant Majorana fermions within the well-established analytical paradigm of Landau-Fermi liquid theory. We find a qualitative and quantitative agreement between the
salient features of {our Landau-Majorana liquid theory and the 
Kitaev spin liquid candidate Ag$_3$LiIr$_2$O$_6$}. Our study 
presents strong evidence for a Fermi liquid-like ground state in the fundamental excitations of a honeycomb iridate, and opens new experimental avenues to detect itinerant Majorana fermions in condensed matter systems.
\end{abstract}

\maketitle

\section{I. Introduction}
Kitaev honeycomb materials 
are among the most promising 
means of realizing itinerant Majorana fermions in a condensed matter setting \cite{Kitaev2006,Hermanns2018Mar,Takagi2019Apr}.
In the quantum spin liquid (QSL) phase, localized spins on the 2D hexagonal lattice fractionalize, allowing us to map the original spin degrees of freedom onto a Hamiltonian describing itinerant fermionic quasiparticles and localized $\mathbb{Z}_2$ fluxes \cite{Savary2016Nov,Do2017,Knolle2019Mar}.
Unfortunately, real materials are more complex than Kitaev's original formulation, and are often characterized by additional interactions 
such as 
an anti-ferromagnetic (AFM) Heisenberg coupling and a symmetric off-diagonal exchange \cite{Rau1,Jackeli2009Jan,Chaloupka2010,Jiang2011Jun,Reuther2011Sep,Price2012Nov,PhysRevB.93.174425,Winter2016Jun}. 
Due to the ubiquitous nature of these non-Kitaev interactions in real-world honeycomb materials, the full effect of off-diagonal contributions to the underlying Majorana excitations is an important question in the study of topological matter. {Recent work on $\alpha$-RuCl$_3$} suggests that the combination of zigzag AFM order and off-diagonal exchange interactions can drastically change the Majorana fermion's band structure {in a nearby spin liquid phase}, ultimately transforming the original Kitaev model's Majorana-Dirac point at zero energy into a Majorana-Fermi surface of finite area \cite{Takikawa2019}. 
As a consequence, such a Majorana-Fermi surface might be induced in a Kitaev material whenever spins exhibiting AFM order are coupled to those in the QSL phase via some exchange interaction. 


Inspired by {previous} works on Majorana-Fermi surfaces \cite{Baskaran1,Tikhonov1,Hermanns1, Hermanns2,Zhang1,Yao1,Chua1,Lai1,Chari1,Erten2017Aug}, in this {paper} we combine theoretical and experimental techniques to address whether or not the specific heat data of { Ag$_3$LiIr$_2$O$_6$} suggests the presence of a Majorana-Fermi surface. { Ag$_3$LiIr$_2$O$_6$} has recently been proposed as a promising spin liquid candidate \cite{Faranak,PhysRevB.103.214405,Tafti2021}, with an enhanced trigonal distortion in the lattice of this compound \cite{SUPP} supporting proximity to the Kitaev QSL phase due to considerable off-diagonal exchange interactions \cite{Katukuri2015Oct,Yadav2016Nov,Haraguchi2020,Tafti2021}.
In a similar vein, a comparative study of the magnetic susceptibility, heat capacity,
and muon spin relaxation of Ag$_3$LiIr$_2$O$_6$ hints at some form of incommensurate AFM order \cite{Faranak2}, a phase previously suggested to {exist in close proximity to} the spin liquid state and dependent on a finite off-diagonal exchange term in the Kitaev Hamiltonian \cite{Rau1}. In relation to {previous theoretical work on the off-diagonal exchange in candidate Kitaev materials}\cite{Takikawa2019}, this raises the question of 
whether or not the exotic thermal signatures found in { Ag$_3$LiIr$_2$O$_6$} \cite{Faranak} { are consistent with the presence of a}
Majorana-Fermi surface and for massive Majorana excitations themselves. 
We answer this question in the positive by analyzing {three aspects of the low-temperature specific heat data:} i) the Sommerfeld coefficient, ii) the next-to-leading-order temperature dependence of the specific heat at zero magnetic field, 
and iii) the low-temperature behavior of the specific heat as a function of the external magnetic field. {This is done theoretically by virtue of a modification of the Pethick-Carneiro calculation of the non-analytic contribution to the Fermi liquid specific heat \cite{Pethick1,Pethick2}, which we extend to the case of a robust Majorana-Fermi surface first introduced in the work of Heath and Bedell \cite{Heath1,Heath2}. Experimentally, we build upon previous work done on the Kitaev magnet candidate Ag$_3$LiIr$_2$O$_6$ by Bahrami et. al. \cite{Faranak}, which is synthesized from the parent compound $\alpha$-Li$_2$IrO$_3$.}
 { The experimental parameters we extract from our analysis show an interdependence in good agreement with the Fermi liquid predictions. {In particular, we find i) a finite Sommerfeld coefficient in the absence of an external magnetic field, ii) leading-order quadratic-$T$ dependence in the specific heat, and iii) magnetic-field dependence of the specific heat which all may be explained in the context of the Landau-Majorana liquid theory \cite{Heath1,Heath2}.}
{These findings are significant, especially in the context of recent resonant inelastic x-ray scattering (RIXS) experiments, where a continuum of magnetic excitations has been found well above the N\'{e}el temperature ($T_N$) {in $\alpha$-Li$_2$IrO$_3$ \cite{Ravelli}, with similar collective magnetic excitations observed in Ag$_3$LiIr$_2$O$_6$ \cite{Tafti2021}}.
%
%
%
%
%
Since well-defined magnons cannot exist at $T>T_N$, the RIXS results have been interpreted as spin-spin correlations within the Kitaev model~\cite{Ravelli}.
Our experimental results for 
$T\ll T_N$ complements and expands upon the conclusions of the RIXS experiments, as we will illustrate that our specific heat data is consistent with a Landau-Majorana liquid and is inconsistent with a dominant magnonic contribution.
%
As such, the combined theoretical and experimental analysis presented here could assist in the selection of potential candidate systems for further investigation and detection of Majorana excitations.
}
\\\\
\section{II. Results and Discussion}
\vspace{2mm}
\subsection*{\normalsize II A. Theory of the Majorana-Fermi surface}
The many-body Majorana model we consider is built off of a modification of fermionic combinatorics, where self-annihilation results in a modulo-2 correction to the traditional "stars and bars" argument of the statistical weight \cite{Heath1}. At low temperatures, the resulting distribution function 
resembles a Fermi-Dirac distribution, albeit with a much more sharply-defined discontinuity in the thermodynamic limit. {More specifically, the "Majorana-Fermi surface" discussed in \cite{Heath1} retains a strong step function-like behavior in momentum space even at finite temperature, whereas the traditional Fermi surface of complex Dirac fermions is characterized by some finite "smearing" as we raise the temperature.} This sharp Majorana-Fermi surface has been found to be a universal many-body feature of independent self-conjugate particles, and as a result 
the Fermi-Dirac Sommerfeld coefficient is modified by a multiplicative factor dependent on the dimensionality \cite{Heath1}.
At $T=0$, the ground state of the Majorana-like system is a filled Fermi sea and, hence, the Fermi energy $\epsilon_F$ and the Fermi temperature $T_F$ are identical to that of a weakly correlated Fermi liquid. 
Nevertheless, the sharp Majorana-Fermi surface results in a suppression of quasihole excitations \cite{Heath2}, and thus we expect important deviations from traditional Fermi liquid behavior.

In a conventional Fermi liquid, correlations between quasiparticles and quasiholes at finite temperature result in 
long-wavelength fluctuations of the statistical quasiparticle energies \cite{Pethick1,Baym_book}. Here, "statistical" quasiparticle energies obey the same equations as the quasiparticle energies defined by Landau, and are therefore different from the "dynamical" quasiparticle energies corresponding to the poles of the propagator. Long-wavelength fluctuations of these energies lead to a non-analytic scattering amplitude, and as a consequence apparent logarithmic divergences emerge in low-temperature quantities of physical interest \cite{Anderson, Pethick1, Pethick2,Andy}.
Such a phenomenon stems from
the total energy contribution of a quasiparticle of momentum $p$ interacting with other quasiparticles of momentum $p+q$, given as $\Delta \epsilon_{p\sigma}=\sum_{q} f_{p\sigma,\,p+q \sigma'}n_{p+q,\,\sigma'}\label{Eqn5}
$,
where $f_{p\sigma,\,p+q}=\partial \epsilon_p/\partial n_{p+q}$ is the Landau parameter and $n_{p+q,\,\sigma}$ is the equilibrium distribution function. 
The underlying $q$-dependence of the Landau parameter results in the non-analytic behavior seen in $\Delta \epsilon_{p\sigma}$ and the specific heat.
From a k-matrix analysis of these scattering events, a Landau parameter $f^\lambda_{p,\,p+q}=f^\lambda(0)+b^\lambda (\hat{p}\cdot \hat{q})^2+...$ is found, where $\lambda=s\,(a)$ for the spin (anti-)symmetric channel and $b^\lambda$ is a function of the $k$-matrix \cite{Pethick1,Baym_book}. 

Theoretically, we find the presence of a sharply-defined Majorana-Fermi surface modifies the $q$-dependence of the Landau parameter \cite{SUPP}, which now goes as
$f^\lambda_{p,\,p+q}=f^\lambda(0)+\textrm{\foreignlanguage{russian}{b}}^\lambda (\hat{p}\cdot \hat{q})+b^\lambda (\hat{p}\cdot \hat{q})^2+...$, with coefficients $b^\lambda$ and $\textrm{\foreignlanguage{russian}{b}}^\lambda$ (Cyrillic b) related to the scattering amplitude. { As seen in the explicit calculation of $b^\lambda$ and $\textrm{\foreignlanguage{russian}{b}}^\lambda$ \cite{SUPP}, both parameters have strong dependence on the value of the "Majorana parameter" $\alpha$,}
which controls how "sharp" the distribution function is at the Fermi energy. {More explicitly and as introduced in \cite{Heath2}, $\alpha$ is a phenomenological parameter in the number distribution which quantifies the affect fermionic self-conjugation has on collective, many-body behavior.} As $\alpha\rightarrow \infty$, we recover the Fermi-Dirac distribution, with 
$\textrm{\foreignlanguage{russian}{B}}^\lambda\rightarrow 0$ and a value of $B^\lambda$ identical to the standard Fermi liquid result \cite{Pethick1}. As $\alpha\rightarrow 0$, {the fermionic number distribution retains a more step-function like behavior, and as such} the Fermi surface becomes more sharply defined against external perturbations{. As a consequence,} terms in the Landau parameter proportional to $\textrm{\foreignlanguage{russian}{b}}^\lambda$ become more prominent. {Intermediate values of $\alpha$ correspond to systems in which the statistical effects of self-conjugation are subjugated either by repulsive interactions, external effects, or other similar behavior.}

{ With the full form of the $q$-dependent Landau parameter derived}, the change in the quasiparticle energy $\Delta \epsilon_{p\sigma}$ from quasiparticle-quasihole scattering in a Majorana liquid can easily be calculated via standard phase space integrals.
Note that the final form of $\Delta \epsilon_{p\sigma}$ is found by only considering interactions between quasiparticles of momentum $p$ within a cutoff $q_c$; i.e., $\mid p-p_F\mid \ll q_c\ll p_F$, thereby restricting ourselves
to a "window" around the Fermi surface where Fermi liquid theory remains valid.
The resulting change to the specific heat is then found to be


\begin{align}
\frac{\Delta C_v}{nT} \approx &\frac{\pi^4 }{1152}w^\lambda \log(2)\frac{k_B^2}{\epsilon_F} \frac{T_{\textrm{cut}}}{T_M}\frac{T}{T_F }(A_0^\lambda)^3\left(
1+\frac{13\pi^2}{20}\frac{T}{T_{\textrm{cut}}}
\right)
\notag\\
&
+
\frac{13\pi^4}{160}w^\lambda\frac{k_B^2}{\epsilon_F}\frac{T^2}{T_F^2} \log \mid \frac{T}{T_{\textrm{cut}}}\mid (A_0^a)^2 \label{eqn5}
\end{align}
where we have defined the cut-off temperature to be $T_{\textrm{cut}}=\hbar v_F q_c/2k_B$, the "Majorana temperature" to be $T_M=\hbar  v_F \alpha/2k_B$, and we focus on a single scattering channel $\lambda$. {The cutoff temperature $T_{\textrm{cut}}$ is equivalent to the cutoff temperature given in \cite{Pethick1} for the traditional Landau-Fermi liquid system, and corresponds to $v_Fq_c$ in temperature units. The Landau-Fermi liquid approach discussed in this article remains applicable only for temperatures $T<T_c$. The first term in Eqn. \eqref{eqn5} is a unique feature in the specific heat explicitly connected to the presence of Majorana-like behavior in the underlying fermionic excitations \cite{Heath1, Heath2}, and is a direct result of particle-hole scattering near the more sharply-defined Majorana-Fermi surface. The Majorana temperature (i.e., the parameter $v_F\alpha$ in units of temperature) quantifies the multiplicative constant defining the first term in Eqn. \eqref{eqn5}. As such,} in the limit of $\alpha\rightarrow \infty$, the {leading-order expression in Eqn. \eqref{eqn5}} disappears, as expected in a conventional Fermi liquid {with a traditional Fermi surface}.
{Note that the specific heat now has} strong dependence on three new parameters: $T_{\textrm{cut}}$, $T_M$, and $A_0^\lambda$. The {leading-order term} in $C_v/T$ is therefore greatly enhanced if $T_M/T_{\textrm{cut}}\ll 1$. {More specifically, we see that the first-order term in Eqn. \eqref{eqn5} dominates over the logarithmic term if ${T^2}/{T_M T_F}>>{T^2}/{T_F^2}\log \mid T/T_{\textrm{cut}}\mid$, or, equivalently, if $T_{\textrm{cut}}\exp(T_F/T_M)>>T$. As such, we see that if $T_M$ is very small, there is a large temperature range where the logarithmic term in Eqn. \eqref{eqn5} is negligible compared to the leading-order term. As such, we will proceed in the next section to fit the zero-field data to a quadratic polynomial, and from the fit extract the parameters of physical interest.}
\subsection*{\normalsize II B. Experimental signatures at zero external field}
We now test our theoretical predictions for a Majorana liquid by analyzing the temperature and magnetic field dependence of the specific heat in the proposed Kitaev magnet Ag$_3$LiIr$_2$O$_6$ \cite{Faranak,Faranak2}. { {Ag$_3$LiIr$_2$O$_6$} was synthesized from the parent compound $\alpha$-Li$_2$IrO$_3$, with the inter-layer Li atoms replaced by Ag atoms. As shown previously \cite{Faranak}, the weaker O-Ag-O bonds results in unique behavior in the magnetic susceptibility and the magnetic entropy{ which are consistent with Ag$_3$LiIr$_2$O$_6$ being closer to the Kitaev limit than its parent compound $\alpha$-Li$_2$IrO$_3$. In particular, our motivation to consider Ag$_3$LiIr$_2$O$_6$ comes from previous specific heat measurements: whereas the low temperature behavior of $C_v/T$ for Ag$_3$LiIr$_2$O$_6$ and $\alpha$-Li$_2$IrO$_3$ are fairly similar, the {parent} compound undergoes an {AFM transition} at 15 K, {whereas long-range order begins to set into Ag$_3$LiIr$_2$O$_6$ at 8 K \cite{Faranak, Faranak2}. As such, the thermodynamic observables of Ag$_3$LiIr$_2$O$_6$ describe the collective behavior of fermionic excitations in a material closer to the Kitaev limit than $\alpha$-Li$_2$IrO$_3$, and therefore we} will primarily focus on { Ag$_3$LiIr$_2$O$_6$} in the present work. Nevertheless, our analysis could just as easily be applied to any potential spin liquid candidate close to the Kitaev limit.  
}

The silver intercalated Kitaev magnet in question was
synthesized via a topochemical exchange reaction as described in \cite{Faranak2}. The quality of our samples 
was confirmed by powder x-ray diffraction, transmission electron microscopy \cite{Faranak2}, and magnetization measurements, {where the short-range ($T_F$) and long-range ($T_N$) transitions characteristic of a Kitaev magnet are clearly observed in the magnetic susceptibility. {As mentioned in \cite{Faranak2}, the peak in the magnetic specific heat at 14 K is due to static magnetism, as opposed to originating from many-body quantum entanglement. Note that the low-temperature peaks observed in $\alpha$-Li$_2$IrO$_3$, Na$_2$IrO$_3$, and $\alpha$-RuCl$_3$ are similarly due to AFM ordering \cite{Faranak,PhysRevB.95.144406,Widmann2019Mar}. The higher-temperature peak may signal the onset of fractionlization, but in our material the sample develops an instability towards the gapped AFM phase at this temperature instead of melting into the spin liquid phase (a phenomenon also observed in $\alpha$-RuCl$_3$ at low external magnetic field \cite{Banerjee2018Feb} ).}
%
%
%
%
The specific heat was measured 
from 80~mK to 4~K using a custom-built cell installed in a dilution refrigerator. Ruthenium oxide thick film resistors were used for thermometry, and the data were collected with a quasiadiabatic technique \cite{Morin1962}. { Note that all data shown in this letter were taken on samples in the "clean limit batch" discussed in \cite{Faranak2}, where disorder originating from silver inclusion within the honeycomb layers is negligible {and a long-range order at 8 K is observed.}}


\begin{figure}[H]
   \adjustbox{minipage=1.3em,valign=t}{\subcaption{}\label{sfig:testb}}%
   \begin{subfigure}[t]{\dimexpr.5\linewidth-1.3em\relax}
\centering
  \includegraphics[width=1\linewidth,valign=t]{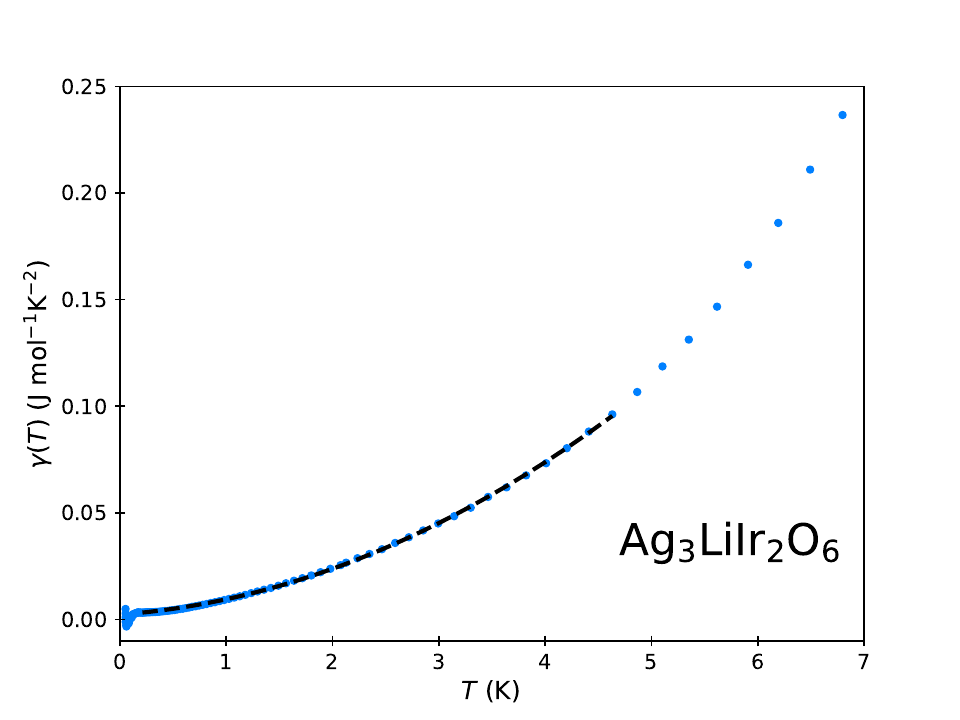}
  \end{subfigure}%
   \adjustbox{minipage=1.3em,valign=t}{\subcaption{}\label{sfig:testb}}%
  \begin{subfigure}[t]{\dimexpr.5\linewidth-1.3em\relax}
  \centering
  \includegraphics[width=1.0\linewidth,valign=t]{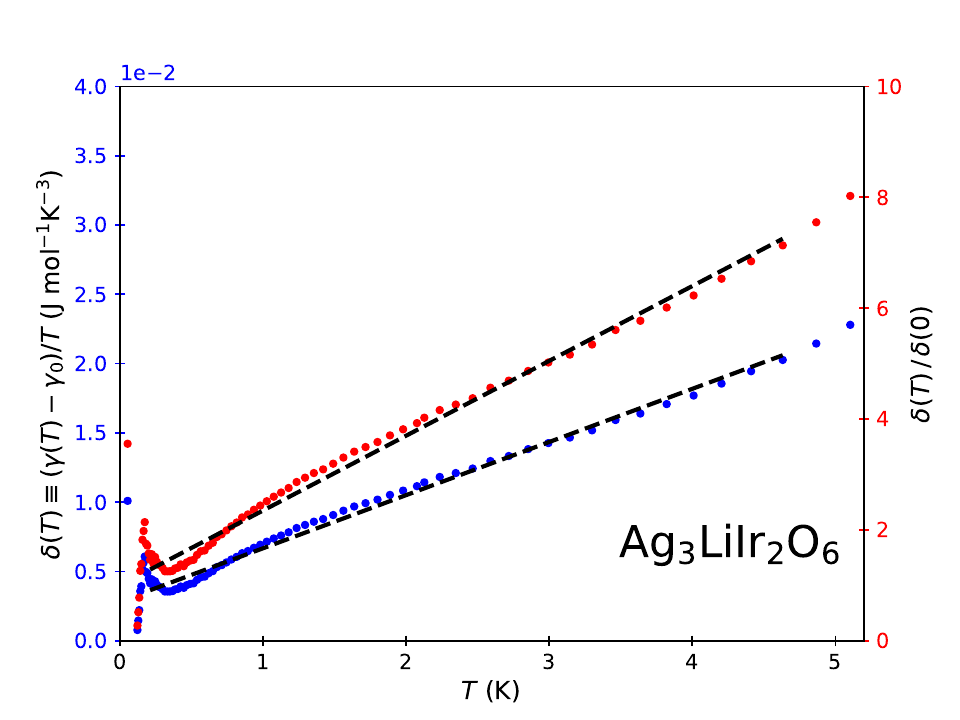}
  \end{subfigure}
  \vspace{5mm}
\\
 \hspace{-5mm} \adjustbox{minipage=1.3em,valign=t}{\subcaption{}\label{sfig:testb}}%
   \begin{subfigure}[t]{\dimexpr.5\linewidth-1.3em\relax}
\centering
  \includegraphics[width=1\linewidth,valign=t]{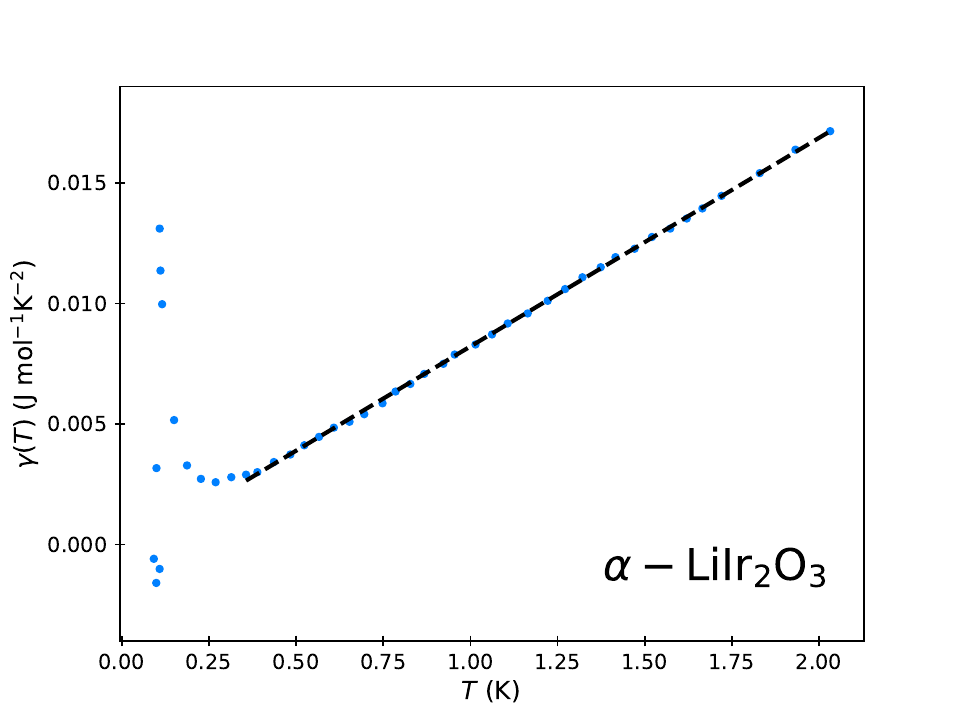}
  \end{subfigure}%
   \adjustbox{minipage=1.3em,valign=t}{\subcaption{}\label{sfig:testb}}%
  \begin{subfigure}[t]{\dimexpr.5\linewidth-1.3em\relax}
  \centering
  \includegraphics[width=1\linewidth,valign=t]{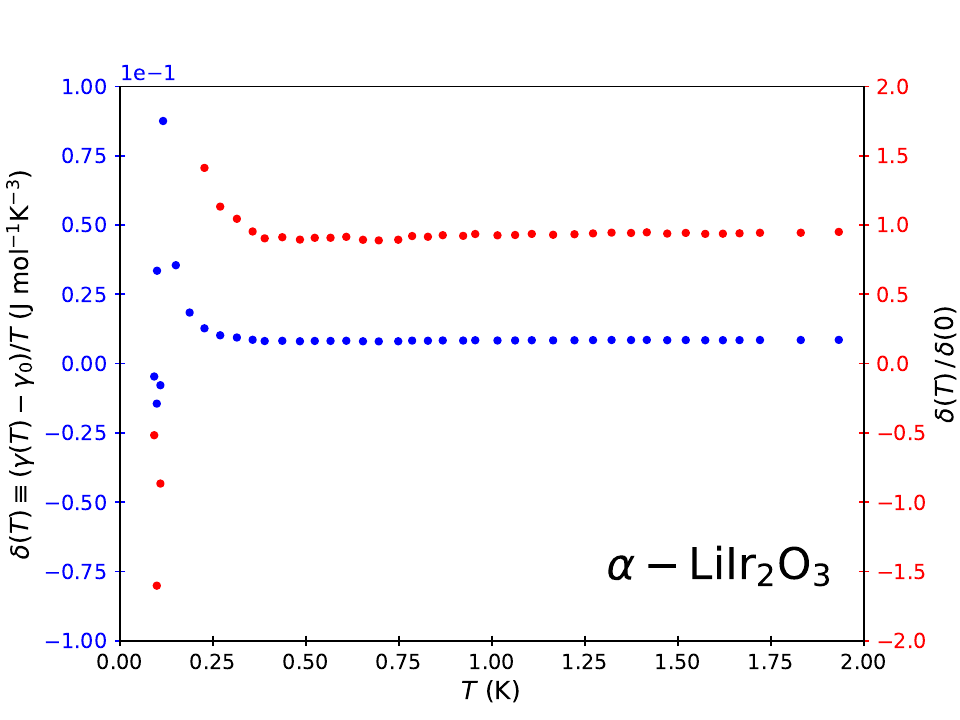}
  \end{subfigure}
\vspace{3mm}
\caption{{\bf Specific heat vs. temperature for Ag$_3$LiIr$_2$O$_6$ and $\alpha-$LiIr$_2$O$_3$.} 
{ 
Measured values of the specific heat $C(T)$ divided by temperature $T$ (denoted by $\gamma(T) \equiv C(T)/T$)  vs $T$ for (a) Ag$_3$LiIr$_2$O$_6$ and (c) $\alpha-$LiIr$_2$O$_3$, both with the Schottky contributions subtracted. We also give the plots of $\delta(T)\equiv (\gamma(T)-\gamma_0)/T$ for the same compounds (shown in (b) and (d), respectively).}
The specific heat data for { $\alpha-$LiIr$_2$O$_3$} is taken from the work of Mehlawat, Thamizhavel, and Singh \cite{PhysRevB.95.144406}. {For plots of the raw data of the Ag$_3$LiIr$_2$O$_6$, see Figs. \ref{fig:6} and \ref{fig:7} in the Methods section.} The data {in (a)} fits to a general quadratic, in agreement with Eqn. \eqref{eqn5}. {However, the general temperature dependence for the specific heat in $\alpha-$LiIr$_2$O$_3$, as shown in (c), lacks the strong quadratic behavior predicted from Eqn. \eqref{eqn5}. The plotted values of $\delta(T)$ in (b) and (d) represent the same specific heat data, but plotted in such a way to discern possible linear-$T$ dependence. The blue dots (left axis) are the { unrenormalized} data of {$\delta(T)$ defined in text}, while the red dots (right axis) are the data renormalized by the projected $T=0$ value. A finite y-intercept in (b) suggests a finite $T/T_{cut}$ term in $\gamma(T)$. 
}}
\label{fig:1}
\end{figure}

{In Figs. \ref{fig:1}(a) and Fig. \ref{fig:1}(b), the quantities $\gamma(T)$ and $\delta(T)\equiv (\gamma(T)-\gamma(T=0))/T$, respectively, are plotted vs. $T$ for our Ag$_3$LiIr$_2$O$_6$ sample. In Figs. \ref{fig:1}(c) and Fig. \ref{fig:1}(d), the same quantities are plotted, except for $\alpha-$LiIr$_2$O$_3$. In Fig. \ref{fig:1}(a),} a finite Sommerfeld coefficient $\gamma(0)\equiv \gamma_0$ is apparent from the {low-temperature data of Ag$_3$LiIr$_2$O$_6$}.
The contribution $\gamma_{Sch}$ from the Schottky anomaly \cite{Schottky,Gopal1966} is subtracted off, with the anomaly modeled as a two-level system with a gap $\Delta$; i.e., $
    \gamma_{Sch}(T)=\sigma T^{-3}  e^{-\Delta/T}/(1+e^{-\Delta/T})^2
$, where $\sigma\sim \Delta^2$ is the Schottky coefficient. As explained in the Methods section, a strong nuclear origin to the Schottky anomaly {does not fit to the model used in our temperature/magnetic-field analysis}. Alternatively, the observed increase of the {low-temperature} specific heat could originate from $\mathbb{Z}_2$ fluxes, which may be modeled by a two-level Schottky-type formula \cite{Tanaka2020Jul}.
With this contribution subtracted, we fit { $\gamma(T)$ as a function of temperature} {for Ag$_3$LiIr$_2$O$_6$} to a quadratic polynomial, finding good agreement above $0.2$~K and up to $5$~K. {Our motivation for taking such a fit comes from the form of Eqn. \ref{eqn5}, where the presence of Majorana-like quasiparticle scattering near the Fermi surface results in a dominant contribution to the specific heat which goes as $T^2/T_MT_F$. Disagreement from the quadratic fit below $0.2$~K in Fig. \ref{fig:1}(a)} {and Fig. \ref{fig:1}(b)}} may be attributed to residual nuclear effects in $\gamma_{Sch}(T)$ at very low temperatures{; namely, the nuclear-specific Schottky anomaly results in the sharp upturn and decline in the low-temperature behavior of $\gamma(T)$}. From the fit, the Sommerfeld coefficient is estimated to be $\approx 2.5$~mJ mol$^{-1}$ K$^{-2}$. {Note that, if we take} the inter-particle distance {(i.e., the Ir-Ir distance in the honeycomb layer)} in the {Ag$_3$LiIr$_2$O$_6$} {sample to be $a=3.03$} \AA\,, {the mass of the fermionic-like excitations is found to be within the same order of magnitude as the bare electron mass. The relative comparability of the bare electron mass to the emergent mass of the fermionic quasiparticles in {Ag$_3$LiIr$_2$O$_6$} may be seen as evidence of a weakly-correlated Fermi liquid-like phase in Ag$_3$LiIr$_2$O$_6$. The breakdown of nearly massless, Dirac fermion-like behavior in said excitations may be seen as a consequence of the off-diagonal exchange interactions in our Ag$_3$LiIr$_2$O$_6$ sample \cite{Takikawa2019}, and may therefore be inferred as evidence that the fermionic quasiparticles in Ag$_3$LiIr$_2$O$_6$ exhibit a quadratic energy dispersion.}

{Our evidence of a finite $T=0$ Sommerfeld coefficient and massive quasiparticles is surprising, as similar Kitaev magnets are generally considered to have a Mott insulating bulk \cite{Trebst2022Mar}. Nevertheless, while the above analysis suggests that Ag$_3$LiIr$_2$O$_6$ may host massive fermionic excitations of some kind in the bulk}, 
we cannot decipher the possible presence (or lack thereof) of {\it self-conjugate} fermionic excitations from the Sommerfeld coefficient alone.
{As such,} we plot $\delta(T)\equiv (\gamma(T)-\gamma_0)/T$ {in {Fig. \ref{fig:1}(b)} and {Fig. \ref{fig:1}(d)} for Ag$_3$LiIr$_2$O$_6$ and $\alpha-$LiIr$_2$O$_3$, respectively}. From Eqn. \eqref{eqn5}, if we ignore the $\log \mid T/T_{\textrm{cut}}\mid$ contribution and the $T$-dependent terms from the non-interacting specific heat, the value of $(\gamma(T)-\gamma_0)/T$ should result in a linear function, with a y-intercept $\propto T_{\textrm{cut}} T_M^{-1} T_F^{-1}$ and a slope $\propto T_{\textrm{cut}}^{-1}$ after renormalization by the former. The plot in Fig. \ref{fig:1}(b) indeed shows clear linear behavior with a finite y-intercept, the latter of which agreeing with our Majorana liquid model. {We emphasize here that the linear$-T$ behavior seen in $C_v/T$ after we subtract the Sommerfeld coefficient is the hallmark of a sharply-defined Fermi surface at finite temperature, and is therefore a telltale signature of the Majorana liquid phase proposed in \cite{Heath1}. A Landau-Fermi liquid would not have this linear-$T$ dependence, but instead would be approximately a constant in $T$. }

From the y-intercept of $\delta(T)$ we find a value of the cutoff temperature, $T_{\textrm{cut}}$, on the order of $2-5$~K. 
An exceedingly small cutoff temperature is expected in the itinerant Majorana theory we consider {here}, as a suppression of hole-like excitations near the Majorana-Fermi surface should severely reduce the difference between the momenta of a quasiparticle and its quasihole neighbors, effectively "narrowing" the Fermi liquid regime  about the Fermi momentum \cite{Andy}. From the renormalized slope, the value of the Majorana temperature may be approximated by $T_M\sim \left(A_0^\lambda
\right)^3 10^{-3} K$.
For strong repulsive interactions, the Majorana temperature saturates to a value much lower than the cutoff temperature, while attractive interactions lead to a breakdown of the underlying theory (both of which are expected \cite{Heath1,Heath2}).

\begin{figure}
  \centering 
  \includegraphics[width=0.75\columnwidth]{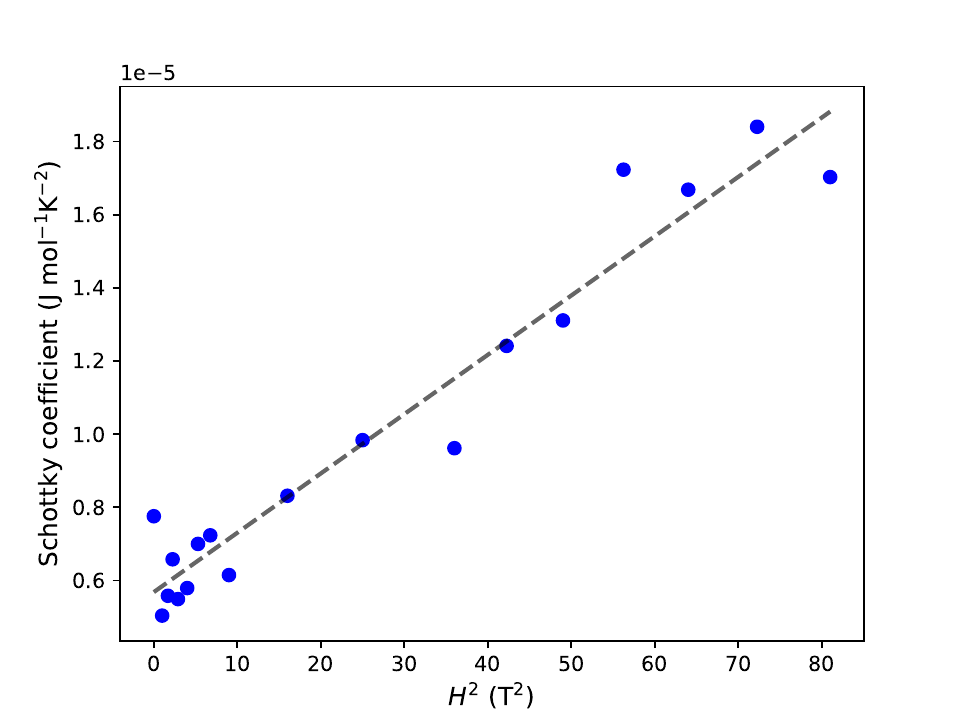}
  \caption{{ \bf Schottky coefficient vs. external magnetic field.}
  Magnetic field dependency of the Schottky parameter, found by fitting to the experimental specific heat data. Blue dots are raw experimental data, while the dotted line is the linear fit {to the raw data}. Above an external field $H=4$~T, linear behavior suggests a gap $\Delta$ of the $\mathbb{Z}_2$ fluxes which grows {proportional to} $H$. As the Majorana gap $\Delta_M$ should go as $H_x H_y H_z/\Delta^2$ in extended Kitaev-Heisenberg materials at high fields \cite{Takikawa2019}, this would suggest that $\Delta_M\sim H$ in our case, in agreement with related Kitaev-like materials such as $\alpha$-RuCl${}_3$ \cite{PhysRevLett.119.037201,PhysRevLett.120.117204}.
  }
  \label{fig:2}
\end{figure}

Before we precede, it is important to perform the above analysis on $\alpha-$LiIr$_2$O$_3$. In Fig. \ref{fig:1}(b), we plot the specific heat $\gamma(T)\equiv C(T)/T$ vs. $T$ and $\delta(T)\equiv (\gamma(T)-\gamma_0)/T$ vs. $T$, respectively, for $\alpha-$LiIr$_2$O$_3$ \cite{PhysRevB.95.144406}. {The specific heat data of $\alpha-$LiIr$_2$O$_3$ reveals a nearly vanishing (i.e., below the mJ range) Sommerfeld coefficient, a dominant linear$-T$ dependence in $\gamma(T)$, and {the lack of any temperature-dependence in $\delta(T)$}. {Compare this to the features seen in Ag$_3$LiIr$_2$O$_6$, where the Sommerfeld coefficient is finite, $\gamma(T)$ is strongly quadratic, and $\delta(T)$ exhibits near-linear $T$ dependence.}
}
%
%
%
%
From our analysis, this would suggest that the Majorana temperature in the {parent} compound is much larger than that of {Ag$_3$LiIr$_2$O$_6$}, which in turn implies that {Ag$_3$LiIr$_2$O$_6$} is closer to the Kitaev limit than its parent compound. This conclusion agrees with previous results on these two iridate materials, where it was illustrated that {Ag$_3$LiIr$_2$O$_6$} is closer to the Kitaev limit than $\alpha-$LiIr$_2$O$_3$ \cite{Faranak, Faranak2}. {Similarly, the vanishing Sommerfeld coefficient seen in $\alpha-$LiIr$_2$O$_3$ agrees with the analysis of Mehlawat et. al. \cite{PhysRevB.95.144406}, and suggests that exchange interaction terms in $\alpha-$LiIr$_2$O$_3$ are not substantial enough to open a gap in the dispersion and subsequently lead to the formation of a Fermi surface \cite{Takikawa2019}.}
{As a consequence, we emphasize that the data shown in Fig. \ref{fig:1} is a clear signature of massive, Majorana-like fermionic excitations in {Ag$_3$LiIr$_2$O$_6$ in the limit of zero external magnetic field.}
}

{Note that the leading-order linear-$T$ dependence of $\gamma(T)$ for $\alpha-$LiIr$_2$O$_3$ 
suggests that this material may host intra-planar complex fermions with a near-linear dispersion, although future research on this material would need to be done to confirm this.  Finally, we emphasize that} the results given above, {for both Ag$_3$LiIr$_2$O$_6$ and $\alpha-$LiIr$_2$O$_3$}, support our claim that the analysis presented in this paper provides a general framework to characterize potential Kitaev spin liquid candidates through the lens of specific heat measurements, and is not specialized to just Ag$_3$LiIr$_2$O$_6$.
%
%
%
}
\subsection*{\normalsize II C. Experimental signatures at finite external field}
Additional experimental signatures of a Majorana-Fermi surface can be {extracted} by turning on a finite magnetic field. To measure specific heat,
thermometry was 
calibrated as a function of temperature in a magnetic field $H$ up to $9$~T.
In Fig. \ref{fig:2}, we plot the Schottky coefficient {as a function of} $H^2$, revealing clear linear behavior above $H=5-6$~T. From our two-level Schottky formula, this behavior suggests that the gap $\Delta$ of the $\mathbb{Z}_2$ fluxes is linearly dependent on the magnetic field. 
As a consequence, the Majorana gap should also go linear with $H$ when the external field is large \cite{Kitaev2001,Takikawa2019,Tanaka2020Jul}, a result which agrees with recent experiments on the related material $\alpha$-RuCl$_3$ \cite{PhysRevLett.119.037201,PhysRevLett.120.117204}.

In Fig. \ref{fig:3}(a) and Fig. \ref{fig:3}(b), we plot the experimental values of the change in specific heat $\Delta \gamma(H,T) =\gamma(0,\,T)-\gamma(H,\,T)$ {as a function of} $T$. The data  
shows a restricted temperature interval with apparent linear $T$-dependence on the semi-log scale (orange background in Fig. \ref{fig:3}), in addition to non-monotonic $H$-dependence exhibited in  $\gamma(H,\,T)$.
%
We propose that, much as in the case of $H=0$, such unconventional behavior can be explained as a  consequence of some liquid-like phase of itinerant Majorana fermions.
In a regular Fermi liquid, the magnetic field dependence of the specific heat is found by calculating the magnetic susceptibility and exploiting fundamental Maxwell relations \cite{Carneiro1977Sep,Misawa1983Feb}. Utilizing the form of the Majorana liquid's quasiparticle energy we have already derived, it can easily be shown that $\Delta \gamma(H,\,T)\sim -H^2 \log \mid T/T_{\textrm{cut}}\mid$\cite{SUPP}.
As such, we expect $\Delta \gamma(H,\,T)$ to increase logarithmically in temperature before reaching the cutoff {temperature}, $T_{\textrm{cut}}$, after which a Fermi liquid-like description is no longer appropriate. This crossover is clearly observed in Fig \ref{fig:3}, with { $T_{\textrm{cut}}$} marked by a dashed line with the label $T_{\textrm{cut}}$. Directly below $T_{\textrm{cut}}$, the experimental data confirms the emergence of a logarithmic temperature dependence as predicted by our theory. {Similarly, in Fig. \ref{fig:4_mag_new}, we see the general behavior of $\Delta \gamma(H,T)$ roughly follows quadratic dependence in the magnetic field $H$ for a finite temperature range below $T_{cut}$, which provides further evidence of a Fermi liquid-like phase at finite magnetic field.}

\begin{figure}[H]
\hspace{0mm}\adjustbox{minipage=1.3em,valign=t}{\subcaption{}\label{sfig:testa}}%
  \begin{subfigure}[t]{\dimexpr.5\linewidth-1.3em\relax}
\centering
  \includegraphics[width=1.6\linewidth,valign=t]{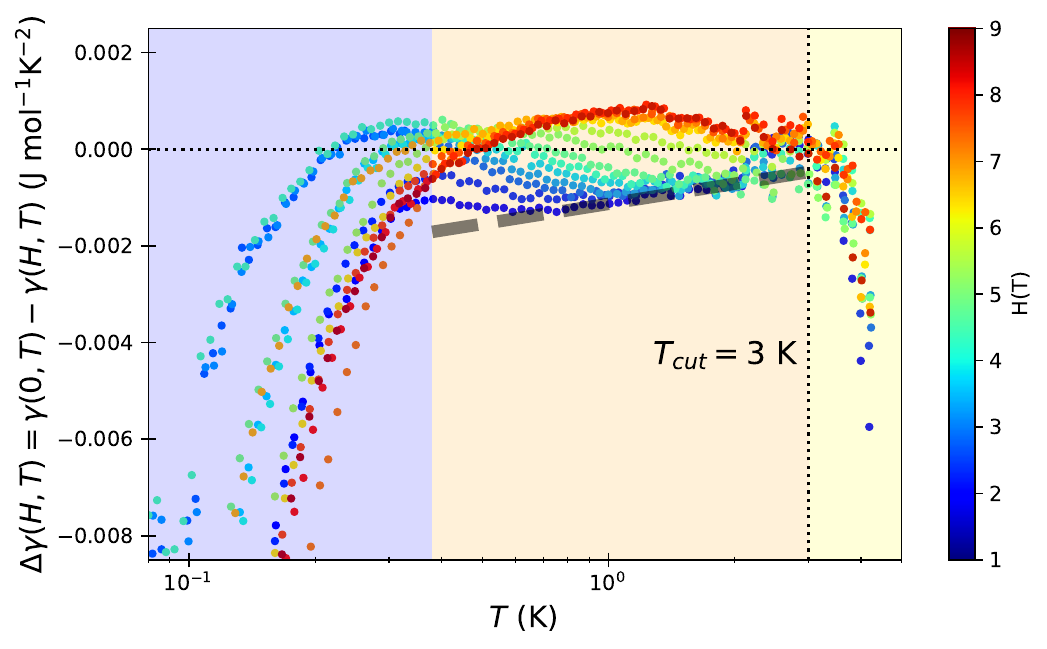}
  \end{subfigure}%
  \hspace{-5mm}
\\\\  
\hspace{-5mm} \adjustbox{minipage=1.3em,valign=t}{\subcaption{}\label{sfig:testb}}%
  \begin{subfigure}[t]{\dimexpr.5\linewidth-1.3em\relax}
  \centering
  \includegraphics[width=1.6\linewidth,valign=t]{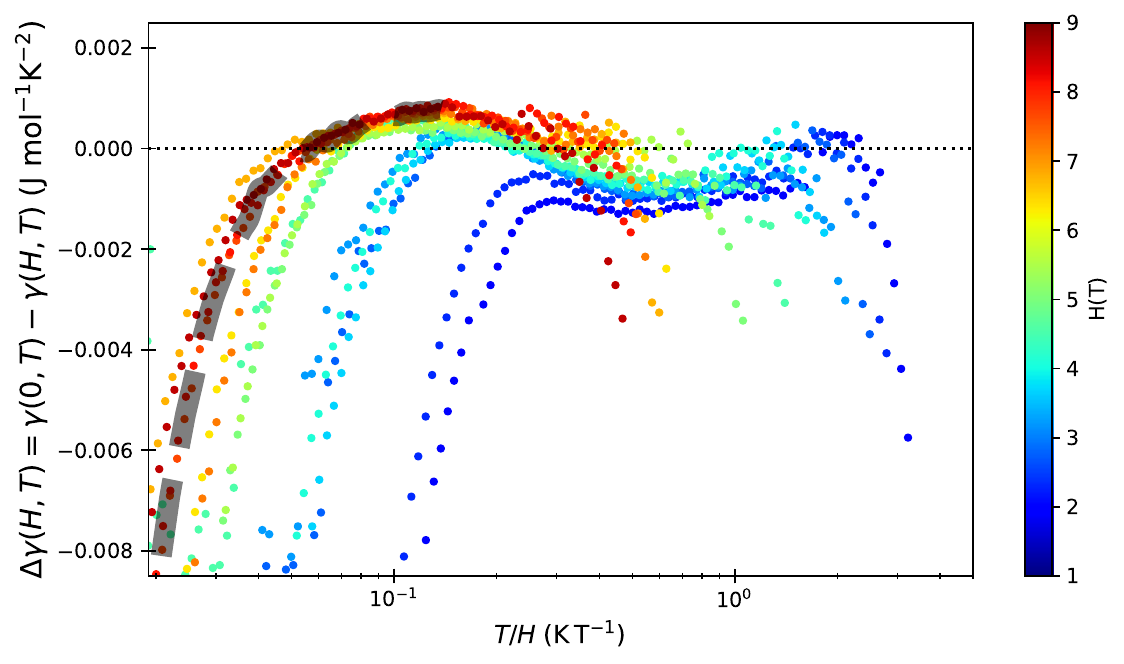}
  \end{subfigure}
  \vspace{3mm}
   \caption{{\bf Change in specific heat vs. temperature at finite magnetic field.} (a) The temperature dependence of $\Delta \gamma(H,T) =\gamma(0,\,T)-\gamma(H,\,T)$ at various values of the external magnetic field $H$, where $\gamma(H,\,T)\equiv C(H,\,T)/T$ denotes the specific heat at finite magnetic field divided by temperature $T$. Within an intermediate temperature regime (orange background), the experimental data at low $H$ follows a straight line on the semi-log scale before sharply decreasing around $T_{\textrm{cut}}=3$~K (vertical dotted line). {The dashed grey line is used as a guide to the eye for the proposed near-linear trend in the data for this region.} Above $T_{\textrm{cut}}$, the logarithmic behavior breaks down (yellow background), as predicted by our itinerant Majorana model. Suppression of $\Delta \gamma(H,\,T)$ at low magnetic field (blue background) may be attributed to the effects of strong correlation.
(b) At high values of $H$, rescaling the temperature with inverse field yields behavior reminiscent of the scaling features seen in H$_3$LiIr$_2$O$_6$ \cite{Kitagawa2018Feb}, where the low-$T$ data approaches the same general curve for high values of the magnetic field, {with the grey dashed curve serving as a guide to the eye for this data collapse}. Experiments on the silver lithium iridate at higher values of $H$ are needed for more conclusive evidence of an eventual collapse to a single scaling curve in this compound.
}
  \label{fig:3}
\end{figure}

\begin{figure}
  \centering 
  \includegraphics[width=0.89\columnwidth]{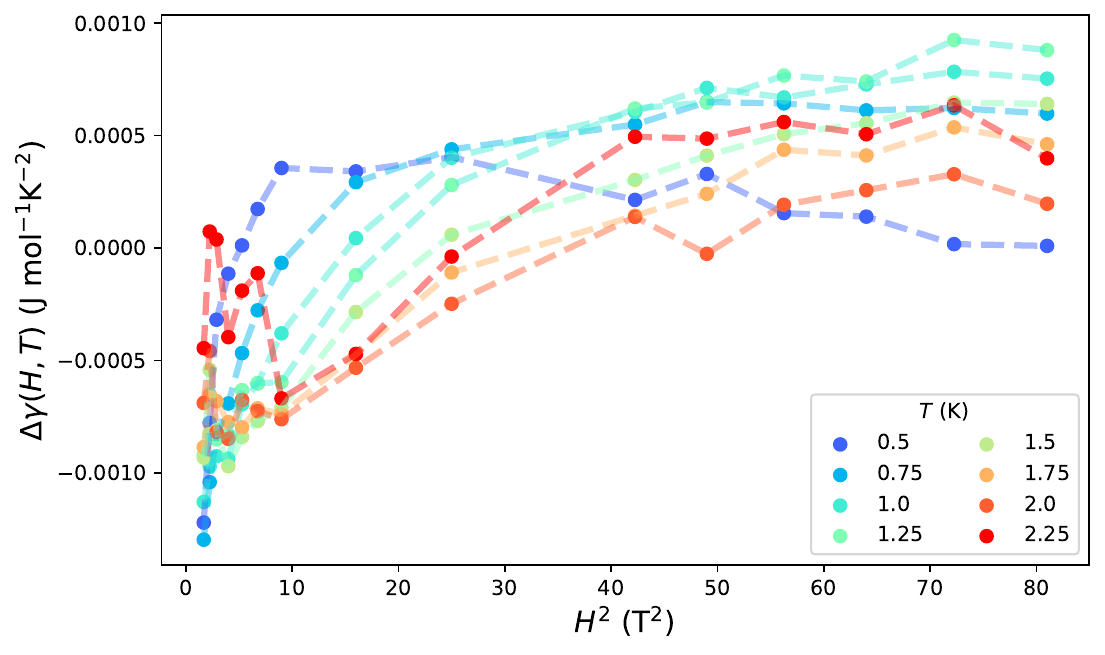}
  \caption{{\bf Change in specific heat vs. external magnetic field.}
{Magnetic field dependency of $\Delta \gamma(H,T)$ for several temperatures, where $\gamma(H,\,T)\equiv C(H,\,T)/T$ denotes the specific heat at finite external magnetic field $H$ divided by temperature $T$. For low temperatures, we have a very fast suppression of $\Delta \gamma(H,T)$ as we increase the external magnetic field. However, as we increase the temperature, we see the data approach a near-linear function of $H^2$. This agrees with the prediction made in the text and derived in this article, where we propose $\Delta \gamma(H,\,T)\sim -H^2 \log \mid T/T_{cut}\mid$. 
Deviation from the linear trend at small magnetic fields and high temperature (seen for $T=2.25$ K and $H\lessapprox 3$~T) may be attributed to the breakdown of a dominant quasiparticle-contribution to the specific heat at weak magnetic fields as we approach $T_{cut}$ from below.
}
  }
  \label{fig:4_mag_new}
\end{figure}

{In Fig. \ref{fig:3}}, the temperature at which logarithmic growth {in $\Delta \gamma(H,\,T)$} appears to vanish is on the order of $3$~K, which is of the same scale as the cutoff temperature estimated from the zero-field data. {Such agreement between the zero and finite field estimates of $T_{\textrm{cut}}$ also suggests a negligible contribution to the specific heat from magnons or phonons, which is inline with previous work on related spin liquid candidates \cite{PhysRevB.99.054426,Majumder,Ravelli}.
{The breakdown of linear behavior on the semi-log scale for temperatures above $0.4$~K (orange background in Fig. \ref{fig:3}) for increasing magnetic field strengths may be attributed to strong magnetic field-induced interactions which dominate over the logarithmic temperature dependence in the weakly-interacting Fermi liquid approximation.}

Turning to Fig. \ref{fig:4}, the behavior of $\gamma(H,\,T)$ as a function of the external magnetic field can also be seen as evidence of a Fermi liquid-like phase. The initial rise of $\gamma(H,\,T)$ with increasing $H$ is often seen in Fermi liquids
in close proximity to a field-induced quantum critical point, where the effective mass is enhanced and consequently the Fermi temperature is lowered \cite{Doniach1966Oct,Limelette2010Jul}. The subsequent fall of the specific heat at larger values of the magnetic field
may then be attributed to the non-analytic behavior in $\gamma(H,\,T)$, which becomes enhanced as 
$H$ is increased \cite{Misawa1983Feb}. 
The low-temperature behavior of $\gamma(H,\,T)$ as a function of external field is therefore in good 
agreement with the predictions of a liquid-like phase of fermionic excitations, with the data in Fig. \ref{fig:4} showing similar trends to a related analysis done on the proposed local Fermi liquid phase of the layered cobalt oxide $[$BiBa${}_{0.66}$K${}_{0.36}$O${}_2]$CoO${}_2$ \cite{Limelette2010Jul}. {Finally, note that, as we raise the temperature, the sharp decline of $\gamma(H,T)$ with increasing magnetic field strength becomes less appreciable. As a consequence, the non-analytic behavior in $\gamma(H,T)$ becomes sub-dominant for higher temperatures, which, as previously mentioned, may explain the breakdown of linear behavior in the plot of $\Delta \gamma(H,T)$ vs. $T$ on the semi-log scale of Fig. \ref{fig:3}a.}

\hspace{0mm}
\begin{figure}
  \centering
 \hspace{0mm}\includegraphics[scale=0.8]{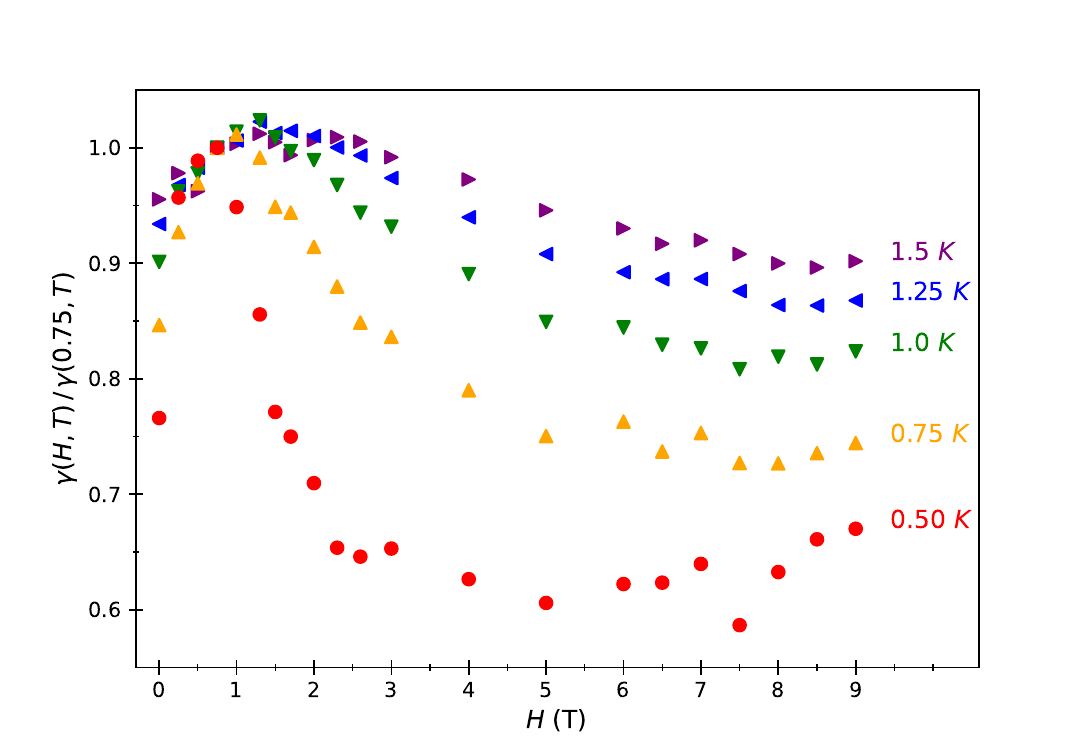}
  \caption{{\bf Specific heat vs. external magnetic field.} Magnetic field dependence of $\gamma(H,\,T)\equiv C(H,T)/T$, where $C(H,\,T)$ is the magnetic-field dependent specific heat $C(H,\,T)$ divided by temperature $T$ and $H$ is the external magnetic field, with data renormalized so the maximum is near unity. As the temperature is lowered, the non-monotonic behavior of $\gamma(H,\,T)$ becomes more pronounced. Temperatures are taken such that $T<T_{\textrm{cut}}$.
  }
  \label{fig:4}
\end{figure}

{ 
%
\subsection*{\normalsize II D. Possible magnon contribution to the specific heat}
{ Before we precede, it is important to bring to light the possible issue of a magnonic contribution to the specific heat measurements. Recent powder inelastic neutron scattering data of the parent compound $\alpha$-Li$_2$IrO$_3$ suggest magnonic excitations at a few meV \cite{PhysRevLett.108.127204,PhysRevB.99.054426}, however direct detection of magnons in the lithium iridates is often challenging as Ir is a strong neutron absorber. Recent RIXS data on the parent compound, $\alpha$-Li$_2$IrO$_3$, have shown a broad continuum of magnetic excitations that persist to 90~K, well above $T_N=15$~K \cite{Ravelli}, with similar magnetic excitations observed in Ag$_3$LiIr$_2$O$_6$ \cite{Tafti2021}. Since well-defined magnons do not exist at $T>T_N$, the RIXS experiments suggest that exotic spin-spin correlations within the Kitaev model are responsible for the continuum observed at high energies (above 10~meV). 
Our results may be interpreted as indirect evidence for a sub-dominant magnon contribution to the specific heat
 in the low-temperature regime $T<T_N$, as a finite Sommerfeld coefficient cannot be explained by magnonic excitations and 
a linear $T$-dependence in $\gamma(T)$ would imply an unconventional magnon dispersion of $\omega\sim k^{3/2}$ unfounded in the iridates \cite{Vladimirov,Pan2013Aug,Winter2017Oct}.

We now turn to the possibility of a magnonic contribution to the specific heat data at high magnetic fields. Theoretically, the magnonic spectrum in an extended Kitaev-Heisenberg system has been studied in the context of nonlinear spin wave theory \cite{PhysRevB.98.060404}, in which case a finite Kitaev exchange ensures a nonzero gap in the magnonic spectrum at high external magnetic fields. The gap, $\Delta$, in the bosonic spectrum would then manifest as an exponential supression $e^{-\Delta/k_B T}$ to the bosonic specific heat. At temperatures on the order of the gap, the gap itself would affect the magnonic dispersion so that $C_V/T\sim T^{1/2}$ with a positive coefficient of proportionality, which is a feature we do not see in our data. As a result, due the ubiquitous nature of a gap in the magnonic spectrum of the Kitaev-Heisenberg model at finite magnetic field, we conclude that the magnonic contribution to our specific heat measurements are negligible compared to the clear Fermi liquid-like signatures seen in Figs. \ref{fig:3} and \ref{fig:4}. Agreement between the zero and finite magnetic field estimates of $T_{\textrm{cut}}$ provide further confirmation for a negligible magnon contribution to the zero-field specific heat data.

Note that our theory of the Landau-Majorana liquid does not completely rule out the presence of magnons in a generic Kitaev magnet. Rather, our theoretical analysis and experimental results suggest that, were magnons leading to a dominant contribution to the specific heat, those magnonic excitations must have very unusual dispersions inconsistent with current theories. In contrast, our data appears to nicely follow the behavior expected from a Landau-Majorana liquid. For future work, it would be helpful to obtain a similarly complete set of temperature and field-dependent specific heat data for other Kitaev spin liquid candidates, and follow a similar analysis as explained here.
%
%
}}
%
%
%
%
%
%
\section{III. Conclusions}
In this letter, the {low-temperature} specific heat $\gamma(H,\,T)$ of {Ag$_3$LiIr$_2$O$_6$} was presented
 at several values of the external magnetic field $H$. A finite Sommerfeld coefficient, in addition to the $T$- and $H$-dependence of $\Delta \gamma(H,\,T)=\gamma(0,\,T)-\gamma(H,\,T)$, {is in agreement with} a Fermi liquid-like ground state in this material. {Likewise, the effective mass of the fermionic-like quasiparticles in Ag$_3$LiIr$_2$O$_6$ have been found to be, rather surprisingly, comparable to the effective mass of the bare electron.} However, the leading order linear-$T$ dependence of $\gamma(0,\,T)$, the linear-$T$ coefficient of $\gamma(0,\,T)$, and the particular breakdown of apparent $T$-logarithmic behavior in $\Delta \gamma(H,\,T)$ all deviate from a traditional Fermi liquid description, and are instead consistent with a Majorana-Fermi surface exhibiting suppressed quasihole excitations \cite{Heath1}. { Similarly, by comparing the specific heat data between Ag$_3$LiIr$_2$O$_6$ and its parent compound $\alpha$-Li$_2$IrO$_3$, our analysis suggests the former is closer to the Kitaev limit than the latter, as was found in previous work \cite{Faranak, Faranak2}.} The high quality of our samples, in addition to the particular $T$-dependence of $\gamma(0,\,T)$ { and recent theoretical work on the magnonic spectrum of the extended Kitaev-Heisenberg model in an external magnetic field \cite{PhysRevB.98.060404}}, { are inconsistent with a contribution from disorder or magnons, at least in the conventional framework.
As the underlying theory makes use of the phenomenological framework of the Landau-Fermi liquid, our analysis is not specific to {Ag$_3$LiIr$_2$O$_6$}, and may be used to discern proximity to the Kitaev limit for a wide range of promising spin liquid candidates, such as the recently proposed Na$_3$Co$_2$SbO$_6$ and Na$_2$Co$_2$TeO$_6$ \cite{PhysRevB.106.014413}.} 
%
%

In terms of future work, further evidence of itinerant Majorana fermions in Ag$_3$LiIr$_2$O$_6$ may be discerned from the low-temperature, high-magnetic field dependence of $\Delta \gamma(H,\,T)$. In related honeycomb materials such as H$_3$LiIr$_2$O$_6$,
the {low-temperature} behavior of the specific heat exhibits a collapse onto a single universal curve when $C(H,\,T)$ and $T$ are appropriately rescaled by $H$ \cite{Helton2007Mar,Sheckelton2012Jun,Mourigal2014Jan,Sheckelton2014Feb,Kitagawa2018Feb}. Such scaling behavior is highly non-trivial, and is often considered the hallmark of a minority of nucleating spins forming random long-range valence bonds \cite{Kitagawa2018Feb,Kimchi2018Oct,Murayama2020Jan}. The pairing of such spins results in a power-law distribution of exchanges \cite{Bhatt1981, Bhatt1982} and is the consequence of a paramagnetic majority in the presence of quenched disorder \cite{Kimchi2018}. In contrast, the Ag$_3$LiIr$_2$O$_6$ sample studied in this work is in the clean limit \cite{Faranak2}, and is hence incompatible with the theoretical framework of \cite{Kimchi2018Oct}. Our data in the right graphic of Fig. \ref{fig:3} confirms that $\Delta \gamma(H,\,T)$ as a function of $T/H$ does not exhibit a clear collapse onto a single scaling curve, 
and instead approaches a limiting behavior at high magnetic fields (highlighted in grey). 
Whether such uniform limiting behavior is a universal feature of a Majorana liquid in the clean limit remains an open question. 

\section{IV. Methods}

\subsection*{\normalsize IV A. Trigonal Distortion in Ag$_{3}$LiIr$_{2}$O$_{6}$ and $\alpha$-Li$_{2}$IrO$_{3}$}
The iridium atoms are octahedrally coordinated with 6 oxygen atoms (IrO$_{6}$) in the honeycomb layers. In the ideal case the O-Ir-O bond angles are 90$^{\circ}$ in an octahedron. However, in real materials a trigonal distortion is commonly observed as a deviation of the bond angles from their ideal values. The degree of trigonal distortion can be quantified by the bond-angle variance~\cite{Haraguchi2020},
\begin{equation}
\sigma = \sqrt{\sum^{12}_{i=1} \frac{(\phi_{i} - \phi_{0})^{2}}{m -1}}
\end{equation}
where $m$ and $\phi_{0}$ are the number and ideal value of O-Ir-O bond angles in an octahedron without distortion, respectively. The values for $\phi_{i}$ are calculated from the Crystallographic Information File (CIF) for both iridate compounds and are presented in Table ~\ref{bond-angle}.

\begin{table}[h]
\caption{\label{bond-angle} The experimental bond-angles for $\alpha$-Li$_{2}$IrO$_{3}$ and Ag$_{3}$LiIr$_{2}$O$_{6}$, where n is the number of angle repetitions.}
 \begin{tabularx}{\linewidth}{c|XXXXXX}
   \hline
   \hline
    Material & \multicolumn{6}{c}{O-Ir-O Bond-Angle }\\
    \hline
$\alpha$-Li$_{2}$IrO$_{3}$ &  85.7$^{\circ}$(n=3) & 86.3$^{\circ}$(n=1) & 89.5$^{\circ}$(n=3) & 90.3$^{\circ}$(n=2) & 93.9$^{\circ}$(n=1) & 94.9$^{\circ}$(n=2) \\
Ag$_{3}$LiIr$_{2}$O$_{6}$ &  82.6$^{\circ}$(n=1) & 83.7$^{\circ}$(n=2) & 83.9$^{\circ}$(n=2) & 85.5$^{\circ}$(n=1) & 95.9$^{\circ}$(n=4) & 96.5$^{\circ}$(n=2)\\
   \hline
   \hline
\end{tabularx}
\end{table}

The bond angle variance ($\sigma$) is 3.46(1)$^{\circ}$ for $\alpha$-Li$_{2}$IrO$_{3}$ and 6.39(1)$^{\circ}$ for Ag$_{3}$LiIr$_{2}$O$_{6}$. Trigonal distortion is twice larger in the Ag-exchanged compound. One can expect an enhancement on the effect of off-diagonal exchange coupling and significant difference between the magnetic behavior of parent and exchanged compounds due to a stronger trigonal distortion in Ag$_{3}$LiIr$_{2}$O$_{6}$, as demonstrated by quantum chemistry calculations \cite{Katukuri2015Oct,Yadav2016Nov}.

\subsection*{\normalsize IV B. Electronic and nuclear Schottky contributions to the specific heat}

The specific heat was obtained via a quasi-adiabatic method. The heater was mounted on one side of the sapphire stage, and the pressed pellet of the polycrystalline sample mounted on the other side with GE varnish. A ruthenium oxide resistance thermometer was glued to the free side of the sample pellet, and the weak link to the bath was attached with GE varnish directly to the sample. A heat pulse was delivered to heat capacity stage at $t_0$, and the temperature of the thermometer was measured as a function of time. The thermometer was directly attached to the sample, and since the heat capacity of the thermometer is negligible in comparison to that of the sample, the heat flowing between the thermometer and the sample is negligible. {The thermometer is therefore in good thermal equilibrium with the sample (its lattice, to be precise), and there is no traditional $\tau_2$ effect. The multi-exponential relaxation of thermometer's temperature is due to internal equilibration processes, as described below.
}

At temperatures below roughly 100 to 150~mK, depending on the magnetic field of the measurement, the raw temperature decay curves display two regimes: initial fast relaxation followed by a very long relaxation, as displayed in Fig. \ref{fig:6}.

The short time scale temperature decay may reflect both the electronic and nuclear degrees of freedom (so called spin-lattice relaxation). We argue below that the nuclear contribution is reflected in the long-time tail of the temperature relaxation curve. At that time, the system (electrons and lattice) is slowly decaying down to the bath temperature. The long-time decay, therefore, reflects the total (electronic plus nuclear) specific heat of the sample, and the short time decay reflects the electronic specific heat only.

The analysis of the zero field data is particularly instructive. The electronic specific heat is determined from the fast decay. The nuclear specific heat can be approximated by the extrapolation of the long time tail of the data back to $t=0$, fitting the last 25\% of the temperature trace to either slow exponential or (perhaps better) linear dependence. A linear fit may be more reliable due to the small part of the temperature curve used for extended extrapolation to $t_0$. Perhaps the strongest indication of the separate contribution from the nuclei is the offset of the short decay exponential fit’s equilibrium temperature from the initial temperature (before heat pulse is applied). This offset is clearly present in the data below roughly 120~mK, as seen in the inset of Fig. \ref{fig:6}.

\begin{figure}[h]
 \hspace{0mm}\includegraphics[width=0.9\columnwidth]{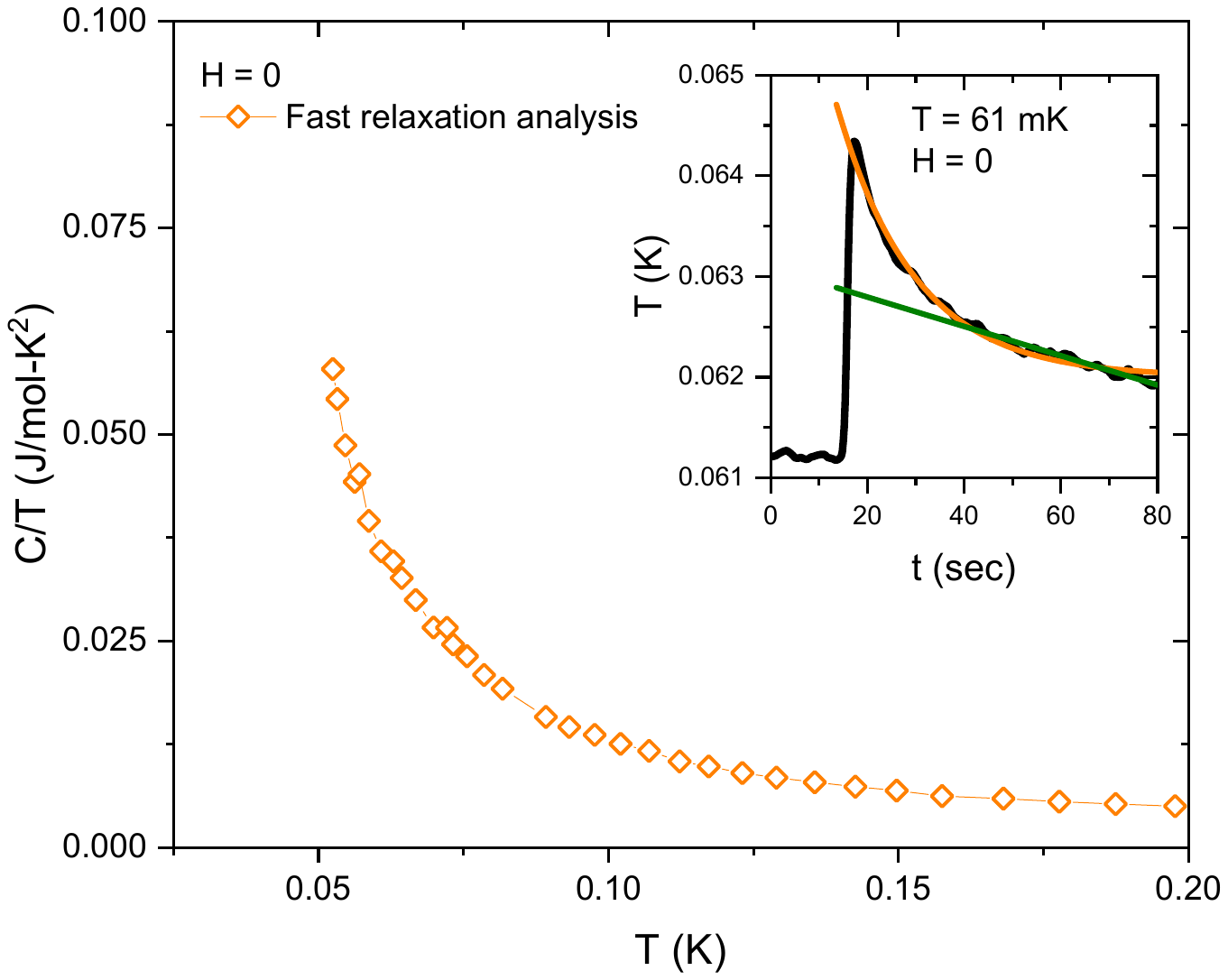} 
\caption{{ Specific heat $C$ divided by temperature $T$ (orange diamonds) obtained from the fast relaxation analysis (20$-$80 sec, e.g. orange curve in the inset). Inset: Temperature decays as a function of temperature for 61~mK in zero magnetic field. Clear offset between the temperature before the heat pulse and the final (equilibrium) temperature of the fast relaxation exponential fit is a reflection of the presence of the
nuclear Schottky anomaly from Ir nuclei. }\phantom{test test test test test test test  test test test test test test test test test test test test test test test test test test}
}
\label{fig:6}
\end{figure}

\begin{figure}[h]
 \hspace{0mm}\includegraphics[width=0.9\columnwidth]{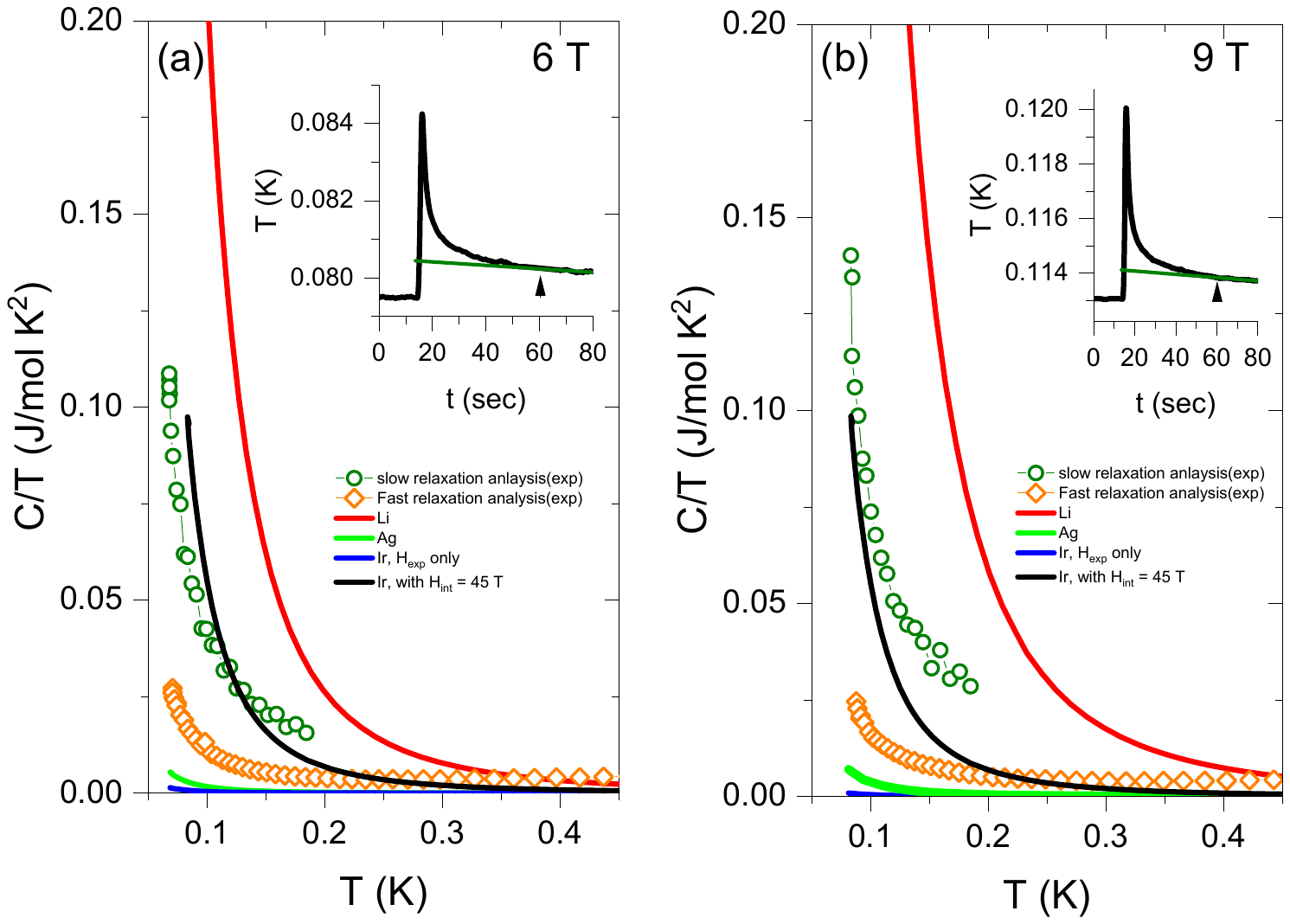} 
\caption{
Specific heat $C$ divided by temperature $T$ obtained from the short time (orange diamonds) and long
time (olive circles) parts of the temperature traces in (a) 6~T and (b) 9~T external field. Schottky anomalies for each nuclei (Li, Ag, Ir) in externally applied magnetic field only are shown by red, green, and blue curves, respectively. Solid black lines represent model calculation of the Schottky anomaly due to Ir nuclei with randomly distributed internal magnetic field ($H_{tot}=\frac{1}{2\pi}\int_0^{2\pi}d\theta \sqrt{
(H_{int}\cos(\theta))^2+(H_{int}\sin\theta+H_{ext})^2}$ due to the polycrystalline nature of the sample, as described in the text. Inset shows an example of the linear analysis of temperature trace in the range of 60$-$80 sec. The missing Li nuclei Schottky anomaly in the data is likely due to very slow relaxation of the Li nuclei spins, similar to that reported for H$_3$LiIr$_2$O$_6$ by \cite{Kitagawa2018Feb}.
}
\label{fig:7}
\end{figure}

The long temperature tail of the temperature decay curves in zero field gives an insight into the origin of the nuclear Schottky anomaly. Generally, it can be either due to the nuclear magnetic spins in the internal magnetic field generated by the magnetically ordered electronic spins, or due to the nuclear electrical quadrupolar moment of nuclei located in positions of non-zero electrical field gradient due to neighboring ions. Li or Ir, both of which have nuclear spin 3/2, have nuclear electrical quadrupolar moment, and potentially may lead to a nuclear quadrupolar anomaly. However, both Li and Ir atoms are situated in high symmetry positions in the crystal lattice (see, for example, \cite{PhysRevB.103.214405}). In accord with this, the quadrupolar splitting of the 7Li NMR line was not observed in a recent study of Ag$_3$LiIr$_2$O$_6$ \cite{PhysRevB.103.214405}. We can safely assume that the same is true for Ir nuclei in the center of the oxygen octahedra cage, i.e. we should expect no quadrupolar Schottky anomaly from Ir nuclei.

The zero field nuclear Schottky contribution must therefore come from the internal magnetic field on Ir nuclei due to the antiferromagnetic order observed to set in at the N\'{e}el temperature $T_N=8$~K \cite{Faranak2}. The results of the analysis of the low temperature data in zero applied magnetic field are displayed in Fig. \ref{fig:6}. The fast decay (orange line in the inset) reflects the electronic degrees of freedom, and the long decay (olive line) is due to the total specific heat, a sum of electronic and nuclear contributions, assuming that a good spin-lattice equilibrium is reached at long times. Electronic contribution can be fitted well by C/T = a/$T^3$, with $a =7.2~J$/$\textrm{mol} \cdot K$. The presence of the two different relaxation time constants in zero field, with the long relaxation time constant due to Ir nuclei in the internal magnetic field produced by the AFM order, proves that the fast relaxation in zero field is entirely of the electronic origin.

We will now consider possible nuclear magnetic moment (spin) contributions in applied magnetic field. Fig. \ref{fig:7} shows the Schottky tail (black solid diamonds) obtained from the analysis of the fast temperature decay. Nuclear magnetic specific heat of Li, Ag, Ir are shown with red, green, and blue symbols, respectively. Contribution from each of the nuclei is calculated as $C_{sch}=Td^2 (RT \log(z))/dT^2$, where $z=\sum g_r \exp(-\epsilon_r/kT)$ \cite{Gopal1966}. Specific heat from Li nuclei greatly overestimates the measured specific heat in external magnetic fields of 6~T and 9~T. This is clearly the result of a very long spin-lattice relaxation rate $1/T_1$ of 7Li, which was observed to crash to zero below the AFM ordering temperature $T_N=8$~K \cite{PhysRevB.103.214405}. Similar low temperature behavior was observed in H$_3$LiIr$_2$O$_6$ \cite{Kitagawa2018Feb}, with 7Li’s spin relaxation rate $1/T_1$ dropping precipitously in a scaling fashion as a function of $T/H$ as a high exponent power law. Thereby Li nuclei are effectively isolated from the electron-phonon lattice, and their specific heat is not measured in our experiment. We therefore rule out any contribution of Li nuclei to the specific heat data resulting from the short decay analysis. Specific heat from Ag nuclei in external magnetic field is rather negligible at 6~T compared to the measured short decay contribution. There is no information at present on the relaxation time behavior of Ag spins. Therefore, Ag nuclei specific heat may contribute to the long relaxation analysis, leading to some uncertainty in the determination of the internal magnetic field on Ir nuclei.

Finally, we can obtain a rough estimate of the internal magnetic field on the Ir nuclei, using the measured electronic (short relation) and total (long relaxation ) specific heat. Ir nuclei contribute to the long temperature relaxation process (see discussion of Fig. \ref{fig:6}). Black solid lines in Figs. \ref{fig:7}(a,b) indicate Schottky anomalies from Ir nuclei in a magnetic field that is a sum of the internal magnetic due to the AFM order and the external magnetic field. Because the sample is polycrystalline, the internal magnetic field due to the AFM order is randomly distributed. We therefore calculated the specific heat due to Ir nuclei with magnetic field $H_{tot}=\frac{1}{2\pi} \int_0^{2\pi} d\theta \sqrt{
(H_{int}\cos(\theta))^2+(H_{int}\sin\theta+H_{ext})^2
}$, where $H_{ext}$ and $H_{int}$ are external and internal magnetic field, respectively. An internal magnetic field of $H_{\textrm{int}}$ = $45\pm 5$ T accounts for data at 6 and 9~T.

We therefore conclude from our analysis that a dominant nuclear origin to the {fast-relaxation} Schottky anomaly in the silver lithium iridate is inconsistent with the temperature and field dependence of the low-temperature data. For this reason, we {interpret} the observed increase of the low-T specific heat as primarily originating from $\mathbb{Z}_2$ fluxes, which may be modeled by a two-level Schottky-type formula \cite{Tanaka2020Jul} as described in the main text. 

\vspace{20mm}

\section{IV C. Non-analytic contributions to the specific heat in a Majorana liquid I: Zero external magnetic field}

In many cases the interactions between quasiparticles and quasiholes complicates a simple "renormalized Fermi gas" picture of the Fermi liquid. A non-analytic contribution to the quasiparticle interaction as the result of some dynamical screening at finite temperature was proposed by Anderson to explain the unusual $T$-dependence in the specific heat of liquid $^3$He \cite{Anderson}. 
This argument was later refined by Carneiro and Pethick \cite{Pethick1,Pethick2} who, in light of previous perturbative calculations, suggested that the low-temperature behavior of the $^3$He specific heat could be explained by long-wavelength fluctuations of the statistical quasiparticle energies. We will use their derivation as a basis for our own study of quasiparticle-quasihole collisions in a Majorana liquid.




{To be more specific, our goal in this appendix is to derive the form of Eqn. 1 in the text. Namely, we wish to show that the values of the Landau expansion coefficients $b^\lambda$ and  $\textrm{\foreignlanguage{russian}{b}}^\lambda$ are given by
\\
\begin{subequations}
\begin{equation}
\textrm{\foreignlanguage{russian}{B}}^\lambda
=\sum_\lambda w_\lambda \left(
\frac{\pi^2 p_F}{8\alpha }(A_0^\lambda)^2 \left[
\frac{1}{ p_F^3 V}-\frac{A_0^\lambda}{36}
\right]
\right)\label{eq2a}
\end{equation}\\ 
\begin{equation}
B^\lambda
=-\sum_\lambda 
\frac{(A_0^\lambda)^2 w_\lambda}{2}\left\{
1-\frac{\pi^2}{12} A_0^\lambda \left(
1-\frac{3\pi^2}{4}\left(\frac{1}{\alpha p_F^2 V}\right)^2
\right\}
\right)\label{eq2b}
\end{equation}
\end{subequations}
\\
\noindent where $V$ is the real-space volume, $w^\lambda$ is the spin multiplicity factor, $A_0^\lambda=F_0^\lambda/(1+F_0^\lambda)$ is the $\ell=0$ scattering amplitude in the $\lambda$-channel, and we take $F_0^\lambda =\nu(0)f_0^\lambda$, $\textrm{\foreignlanguage{russian}{B}}^\lambda\equiv \nu(0) \textrm{\foreignlanguage{russian}{b}}^\lambda$, and $B^\lambda\equiv\nu(0)b^\lambda $, where $\nu(0)$ is the density of states. As discussed in the main text, the "Majorana parameter" $\alpha$ is an experimentally-determined constant which controls how "sharp" the distribution function is at the Fermi energy \cite{Heath2}. 
}

{To derive the form of the $q$-dependent Landau parameter}, we must first consider the scattering of a quasiparticle-quasihole pair in a regular Fermi liquid, the scattering process is dictated by the following equations for the $t$- and $k$- matrices:
\begin{align}
t_{p\,p_1}^\lambda (q,\,\omega)=f_{p\,p_1}^\lambda+2\sum_{p_2}f_{p\,p_2}^\lambda \frac{
n_{p_2}^0-n_{p_2+q}^0
}{
\omega-\epsilon_{p_2+q}+\epsilon_{p_2}+i\eta
}t_{p_2\,p_1}^\lambda (q,\,\omega)
\end{align}
\begin{align}
k_{p\,p_1}^\lambda (q,\,\omega)=f_{p\,p_1}^\lambda+2\mathscr{P}\sum_{p_2}f_{p\,p_2}^\lambda \frac{
n_{p_2}^0-n_{p_2+q}^0
}{
\omega-\epsilon_{p_2+q}+\epsilon_{p_2}
}k_{p_2\,p_1}^\lambda (q,\,\omega)
\end{align}
where $f_{p\,p_2}^\lambda$ is the Landau parameter for the $\lambda$ scattering channel and $n_p^0$ is the equilibrium distribution function.
In the case of particle-particle scattering, the interaction energy is given by the equation

\begin{align}
\Delta E&=K(E)\left\{
1+\sum_{n=1}^\infty \frac{
(-1)^{n}
}{
2n+1
}\left(
\pi \delta(E-H_0)K(E)
\right)^{2n}
\right\}\notag\\
&=K(E)-\frac{\pi^2}{3}K(E)\delta (E-H_0)K(E)\delta(E-H_0)K(E)+...
\end{align}
where $K(E)$ is some general k-matrix which describes mutual particle collisions. For a Fermi liquid, however, we need to consider the quasiparticle-quasihole version of the above, and thus we obtain
\begin{align}
&\Delta \omega_{p\,q}^\lambda\notag\\
&=k_{pp}^\lambda (q,\,\omega_{q\,q})\notag\\
&\phantom{=}-\frac{\pi^2}{3}(2)^2\sum_{p_1 p_2}k_{p\,p_1}^\lambda (q,\,\omega_{p\,q})[n_{p_1}^0-n_{p_1+q}^0]\delta(\omega_{p\,q}-\omega_{p_1\,q})\notag\\
&\phantom{
-\frac{\pi^2}{3}(2)^2\sum_{p_1 p_2}
}\times k_{p_1\,p_2}^\lambda(q,\,\omega_{p\,q})[n_{p_2}^0-n_{p_2+q}^0]\delta(\omega_{p\,q}-\omega_{p_2\,q})k_{p_2\,p}^\lambda(q,\,\omega_{p\,q})+...
\end{align}
where we have defined $\omega_{p\,q}=\epsilon_{p+q}-\epsilon_p$, where $\epsilon_p$ is the quasiparticle energy. 
The interaction energy of the quasiparticle and quasihole singlet spin state is given by $2\Delta \omega_{p\,q}^s$, while the triplet state is given by $2\Delta \omega_{p\,q}^a$. Note that the specific heat is dependent on the spin-symmetric part of the interaction. Given that
\begin{align}
f_{\uparrow \uparrow}(p,\,p+q)=-(\Delta \omega_{p\,q}^s+\Delta \omega_{p\,q}^a)
\end{align}
\begin{align}
f_{\uparrow \downarrow}(p,\,p+q)=-2\omega_{p\,q}^a
\end{align}
we can write the symmetric Landau parameter in the form
\begin{align}
f_{p,\,p+q}^s&=\frac{1}{2}\left(
f_{\uparrow \uparrow}(p,\,p+q)+f_{\uparrow \downarrow}(p,\,p+q)
\right)\notag\\
&=-\frac{1}{2}\left(
\Delta \omega_{pq}^s+3\Delta \omega_{pq}^a
\right)
\end{align}

In the Majorana liquid we consider in the text, the form of the k-matrix differs from that of a traditional Fermi liquid.
To see this, recall from \cite{Heath1, Heath2} the form of the distribution function $\widetilde{n}_{p'}^0$ of the Majorana system:
\begin{align}
\widetilde{n}_{p'}^0=\Theta(p_F-p')+\Theta (p'-p_F) n_{p'}^0
\end{align}
That is, for $p'>p_F$, we have $\widetilde{n}_{p'}^0=n_{p'}^0$, while for $p'<p_F$ we have $n_{p'}^0=1$. As such, we see
\begin{align}
\widetilde{n}_{p'+q}^0&\approx \widetilde{n}_{p'}^0+q\cdot  \frac{\partial \widetilde{n}^0}{\partial p'}
\notag\\
&=\widetilde{n}_{p'}^0+q \cdot \frac{\partial}{\partial p'}\Theta (p_F-p')+n_{p'}^0 q\cdot \frac{\partial}{\partial p'}\Theta(p'-p_F)+\Theta(p'-p_F)q\cdot \frac{\partial n_{p'}^0}{\partial p'}
\end{align} 
We will now take the approximation that $p'$ is slightly above (but very close) to the Fermi surface, eliminating the second term. Now, for the remaining terms, we have

\begin{align}
&n_{p'}^0 q\cdot \frac{\partial}{\partial p'}\Theta(p'-p_F)+\Theta(p'-p_F)q\cdot \frac{\partial n_{p'}^0}{\partial p'}
=n_{p'}^0 |q| \delta(p'-p_F)+\Theta(p'-p_F)q\cdot \frac{\partial n_{p'}^0}{\partial p'}
\end{align}

The first term on the right-hand side of the above is the direct result of the underlying Majorana statistics, which yields a sharp Fermi surface. However, in the case discussed in \cite{Heath1}, we have particle-hole suppression isotropically across the entire Fermi surface, as the smearing from thermal excitations is isotropic (i.e., there is no thermal gradient over $p=p_F$). In this case, the "smearing" is the direct result of a perturbation along the $\hat{q}$-direction; i.e., in the direction of particle-hole propagation. As such, the divergent term we will see is the direct result of an "eternally sharp" point in the distribution along the direction of particle-hole propagation only, and thus the first of the above will be independent of the relative angle between $\hat{q}$ and $\hat{p}$. Hence,
\begin{align}
\widetilde{n}_{p'+q}^0 &\approx 
n_{p'}^0+n_{p'}^0 |q|\delta (p'-p_F)+q\cdot v_{p'}\frac{\partial n_{p'}^0}{\partial \epsilon_{p'}}
\end{align}
and therefore
\begin{align}
\widetilde{n}_{p'}^0-\widetilde{n}_{p'+q}^0&=-n_{p'}^0 |q| \delta(p'-p_F)-q\cdot v_{p'}\frac{\partial n_{p'}^0}{\partial \epsilon_{p'}}
\end{align}
Letting the Heaviside step function be approximated by a generalized Fermi-Dirac-type function $f$, we find that
\begin{align}
n_{p'}^0\delta (p'-p_F) |q|&=n_{p'}^0\frac{f(1-f) |q|}{\alpha}\notag\\
&\approx \frac{|q|}{4\alpha}
\end{align}
Once again, in the above we have assumed that we are close to the Fermi surface. Otherwise, this term disappears. We are left with
\begin{align}
\widetilde{n}^0_{p'}-\widetilde{n}_{p'+q}^0=-\frac{|q|}{4\alpha}-q\cdot v_{p'}\frac{\partial n_{p'}^0}{\partial \epsilon_{p'}}
\end{align}
Performing the relevant phase space integrals, we find that
%
%
%
\begin{align}
\int p'^2 dp'(\widetilde{n}_{p'}^0-\widetilde{n}_{p'+q}^0) \delta ( 
\omega_{pq}-\omega_{p'q}
)
&=\pi^2 \nu(0)\left\{
s-\frac{p_F}{12\alpha} 
\right\} \delta (s-\hat{p}'\cdot \hat{q})
\end{align}

\noindent Hence, for the Majorana liquid, 
\begin{align}
&\nu(0)\Delta \omega_{p\,q}^\lambda\notag\\
&=\kappa^\lambda (\hat{p}\cdot \hat{q},\,\phi=0)
-\frac{\pi^2}{3}\left\{ 
\left(
\frac{1}{2}\hat{p}\cdot \hat{q}
\right)^2
-\frac{ p_F}{48\alpha } \hat{p}\cdot \hat{q}
+\frac{ p_F^2}{576\alpha^2 }
\right\}
\notag\\
&\phantom{-\frac{\pi^2}{3}\bigg\{ }\hspace{10mm}\times \int_0^{2\pi}\frac{d\phi_1}{2\pi}\int_0^{2\pi} \frac{d\phi_2}{2\pi}\kappa^\lambda(\hat{p}\cdot \hat{q},\,\phi_1)\kappa^\lambda(\hat{p}\cdot \hat{q},\,\phi_2)\kappa^\lambda (\hat{p}\cdot \hat{q},\,\phi_1+\phi_2)+...\notag\\
 &=\sum_{m=-\infty}^\infty \left\{
\kappa_m^\lambda (\hat{p}\cdot \hat{q})-\frac{\pi^2}{3}\left\{ 
\left(
\frac{1}{2}\hat{p}\cdot \hat{q}
\right)^2
-\frac{ p_F }{48\alpha }\hat{p}\cdot \hat{q}
+\frac{ p_F^2}{576\alpha^2 }
\right\}[\kappa_m^\lambda (\hat{p}\cdot \hat{q})]^3+...
\right\}
\end{align}
The fact that $\Delta \omega_{pq}$ has linear and quadratic terms in $\hat{p}\cdot \hat{q}$ means that the Landau parameter $F^\lambda_{p,\,p+q}$ will also have linear and quadratic terms in $\hat{p}\cdot \hat{q}$. We can now write
\begin{align}
F^\lambda_{p,\,p+q}=F^\lambda(0)+\textrm{\foreignlanguage{russian}{B}}^\lambda (\hat{p}\cdot \hat{q})+B^\lambda (\hat{p}\cdot \hat{q})^2+...
\end{align}
Our goal now is to find the value of $\textrm{\foreignlanguage{russian}{B}}^\lambda\equiv \nu(0) \textrm{\foreignlanguage{russian}{b}}^\lambda$. This is done by expanding the k-matrices in the expression for $\nu(0)\Delta \omega_{p q}^\lambda$ about $s\equiv \omega/qv_F=0$:
\begin{align}
&\nu(0)\Delta \omega_{p\,q}^\lambda\notag\\
&
=
\sum_{m=-\infty}^\infty \bigg\{
\kappa_m^\lambda (0)+\hat{p}\cdot \hat{q}\frac{d \kappa_m^\lambda}{ds}
\bigg \vert_{s=0}
+\frac{1}{2}\left(\hat{p}\cdot \hat{q}\right)^2 \left[
\frac{d^2\kappa_m^\lambda}{ds^2}\bigg \vert_{s=0}
\right]-\frac{\pi^2}{3}
\bigg\{ 
\left(
\frac{1}{2}\hat{p}\cdot \hat{q}
\right)^2\notag\\
&\phantom{=\sum_{m=-\infty}^\infty \bigg\{}-\frac{p_F}{48\alpha}\hat{p}\cdot \hat{q}
+\frac{p_F^2}{576\alpha^2 |q|^2}
\bigg\}[\kappa_m^\lambda (0)]^3+...
\bigg\}
\end{align}
The term proportional to $(\hat{p}\cdot \hat{q})^2$ is identical to the similar term seen in a regular Landau-Fermi liquid \cite{Pethick1,Pethick2}. The term linear in $\hat{p}\cdot \hat{q}$ is given by
\begin{align}
\textrm{\foreignlanguage{russian}{B}}^\lambda&=-\frac{1}{2} \sum_{\lambda} w_\lambda 
\sum_{m=-\infty}^\infty \left(
\frac{d\widetilde{\kappa}_m^\lambda}{ds}\bigg\vert_{s=0}+\frac{\pi^2}{144}\frac{p_F}{\alpha} [\widetilde{\kappa}_m^\lambda (0)]^3
\right)\notag\\
&=-\frac{\pi^2}{8}\frac{p_F}{\alpha}\left\{
(A_0^s)^2 \left[
\frac{A_0^s}{36}-\frac{1}{p_F^3 V}
\right]+3(A_0^a)^2 \left[
\frac{A_0^a}{36}-\frac{1}{ p_F^3 V}
\right]
\right\}
\end{align}
where we have expanded the k-matrix in terms of partial waves, taken the $\ell=0$ channel as dominant, and used the fact that\footnote{Throughout this derivation, note that we have neglected the additional energy unit and taken $\hbar=1$. }
\begin{align}
\widetilde{\kappa}_0^\lambda (0)=A_0^\lambda
\end{align}
\begin{align}
\frac{d\widetilde{\kappa}_0^\lambda (s)}{ds}\bigg\vert_{s=0}=-(A_0^\lambda )^2 \left\{\frac{\pi^2 }{4\alpha  p_F^2 V}\right\}
\end{align}
\begin{align}
\frac{d^2 \widetilde{\kappa}_0^\lambda(s)}{ds^2} \bigg\vert_{s=0}=2(A_0^\lambda)^2+2(A_0^\lambda)^3 \left(\frac{\pi^2}{4\alpha   p_F^2 V}\right)^2
\end{align}

%
%
Without loss of generality, we can use the same underlying mathematics to calculate $B^s$ for the Majorana liquid:
\begin{align}
B^\lambda
&=
-\frac{1}{2}\left\{
(A_0^s)^2 \left(1-\frac{\pi^2}{12}A_0^s\right)+3(A_0^a)^2 \left(1-\frac{\pi^2}{12}A_0^a\right)
\right\}\notag\\
&\phantom{=-\frac{1}{2}\bigg\{}-\frac{\pi^4}{32}\left(
\frac{1}{\alpha p_F^2 V}
\right)^2 \left\{
\left(A_0^s\right)^3 +3\left(A_0^a\right)^3
\right\}
\end{align}


%
This completes our derivation of the Landau parameter $F_{p,\,p+q}^\lambda =F^\lambda(0)+\textrm{\foreignlanguage{russian}{B}}^\lambda \hat{p}\cdot \hat{q}+B^\lambda (\hat{p}\cdot\hat{q})^2+...$.
So how does this effect the specific heat? We know that the total contribution to the quasiparticle energy coming from $\Delta \epsilon_{p\sigma}$ is given by
\begin{align}
\Delta \epsilon_{p\sigma}=\sum_{q} f_{p\sigma,\,p+q \sigma'}\widetilde{n}_{p+q,\,\sigma'}^0
\end{align}
We will deal with the term linear in $(\hat{p}\cdot \hat{q})$ first. Defining $\textrm{\foreignlanguage{russian}{B}}^\lambda=\nu(0)\textrm{\foreignlanguage{russian}{b}}^\lambda $ and writing $\hat{p}\cdot \hat{q}=\frac{\hat{p}\cdot \vec{q}}{|q|}=\frac{q_{\|}}{\sqrt{q_{\|}^2 + q_{\perp}^2}}$,
the change in the quasiparticle energy from this term is given by
\begin{align}
\Delta \epsilon_p(T)&=\frac{2\textrm{\foreignlanguage{russian}{b}}^\lambda}{(2\pi)^3}\int_{-q_c}^{q_c} dq_{\|} \int_0^{q_c^2-q_{\|}^2} \pi d(q_{\perp}^2)\frac{q_{\|}}{\sqrt{q_{\|}^2+q_{\perp}^2}} \widetilde{n}_{p+q}^0(T)\notag\\
&\approx-\frac{\textrm{\foreignlanguage{russian}{b}}^\lambda}{2\pi^2} q_c\int_{-q_c}^{q_c} dq_{\|} q_{\|}\widetilde{n}^0_{p+q}(T) -\frac{\textrm{\foreignlanguage{russian}{b}}^\lambda}{2\pi^2} \int_{-q_c}^{q_c} dq_{\|}  q_{\|}^2 \widetilde{n}_{p+q}^0 (T)
\end{align}
where we have utilized the fact that 

\begin{align}
\int_0^{q_c^2-q_{\|}^2} d(q_{\perp}^2)\frac{q_{\|}}{\sqrt{q_{\|}^2+q_{\perp}^2}}&=2q_{\|}\sqrt{q_{\|}^2+x}\bigg \vert_{0}^{q_c^2-q_{\|}^2}=2q_{\|}q_c+2q_{\|}^2
\end{align}
We then find the change in the quasiparticle energy is given by
\begin{align}
&\phantom{=}\Delta \epsilon_p(T)\notag\\
&=-q_c\frac{\textrm{\foreignlanguage{russian}{b}}^\lambda}{2\pi^2} \int_{-q_c}^{q_c} dq_{\|} q_{\|} \widetilde{n}^0_{p+q}(T)-\frac{b^\lambda}{2\pi^2}\int_{-q_c}^{q_c} dq_{\|} q_{\|}^2 \log\mid
\frac{q_{\|}}{q_c} 
\mid
\widetilde{n}_{p+q}^0(T)\notag\\
&\phantom{=}-\frac{\textrm{\foreignlanguage{russian}{b}}^\lambda}{2\pi^2}
\int_{-q_c}^{q_c} dq_{\|} q_{\|}^2 \widetilde{n}_{p+q}^0(T)\notag\\ \notag\\
&=\frac{1}{6\pi^2v_F^2} \bigg\{
-\frac{3}{2}(p-p_F)^2v_F^2 \left\{
\frac{3}{2} q_c\textrm{\foreignlanguage{russian}{b}}^\lambda - (p-p_F)\left(
 b^\lambda\log\mid
\frac{\max((p-p_F),\,T)}{q_c}
\mid+  
\textrm{\foreignlanguage{russian}{b}}^\lambda
\right)
\right\}\notag\\
&\phantom{=\frac{1}{6\pi^2 v_F^2}\bigg\{}\hspace{-10mm}+3\log(2) (k_B T)(p-p_F)v_F \left(q_c\textrm{\foreignlanguage{russian}{b}}^\lambda -(p-p_F)\left\{
 b^\lambda\log\mid
\frac{\max((p-p_F),\,T)}{q_c}
\mid+  
\textrm{\foreignlanguage{russian}{b}}^\lambda
\right\}
\right)\notag\\
&\phantom{=\frac{1}{6\pi^2 v_F^2}\bigg\{}\hspace{-10mm}
+\frac{\pi^2}{2}(k_B T)^2 (p-p_F)\left\{
 b^\lambda\log\mid
\frac{\max((p-p_F),\,T)}{q_c}
\mid+  
\textrm{\foreignlanguage{russian}{b}}^\lambda
\right\}\bigg\}
\end{align}
where we found the above by solving Fermi-Dirac type integrals of the form
%
%
\begin{align}
&\phantom{=}\int_{-p+p_F}^{q_c} n_{q+p}' (q+p-p_F)^m dq\notag\\
&=\int_{-p+p_F}^{q_c} \frac{e^{\beta (q+p-p_F)v_F}}{
(e^{\beta (q+p-p_F)v_F}+1)^2
}(\beta (q+p-p_F))^m d q\notag\\
&=-(\beta v_F)^{-m}\int_{0}^{\beta (q_c+p-p_F)v_F} \frac{e^x}{(e^x+1)^2}x^m dx,\quad x=\beta ( q+p-p_F)v_F
\end{align}

We will now derive the change in thermodynamic entropy for the Majorana system. The change in the entropy can be simplified to an integral equation:
\begin{align}
\Delta S&=\sum_p \Delta \epsilon_p(T)\frac{\partial \widetilde{n}_p}{\partial T}
\approx -\frac{p_F^2 }{T\pi^2} \int_0^1 \Delta \epsilon_p(T) \left(p-p_F\right) dn_p
\end{align}
where we have used the fact that
\begin{align}
\frac{\partial \widetilde{n}_p}{\partial T}=\frac{e^{\frac{(p-p_F)v_F}{k_B T}}}{\left(
e^{\frac{(p-p_F)v_F}{k_B T}}+1
\right)^2}\left(\frac{(p-p_F)}{k_B T^2} v_F\right)\Theta(p-p_F)\approx \frac{\partial n_p}{\partial T}
\end{align}
Now, we can run through every term in the above equation for $\Delta \epsilon_p(T)$ and subsequently find the total change in the entropy. These integrals are easy to perform, and are left as an exercise for the reader. The end result is 
\begin{align}
\Delta S&=-\frac{p_F^2}{T\pi^2}\int_0^1 \Delta \epsilon_p(T)(p-p_F)dn_p\notag\\
&=-\frac{13}{240}\pi^4 n k_B \left(\frac{T}{T_F}\right)^3 \left( B^\lambda
\log \vert\frac{T}{T_{\textrm{cut}}}\vert+\textrm{\foreignlanguage{russian}{B}}^\lambda 
\right)\notag\\
&\phantom{=}-\frac{q_c}{8p_F}\log(2) \pi^2 n k_B \textrm{\foreignlanguage{russian}{B}}^\lambda \left(\frac{T}{T_F}\right)^2 
\end{align}
The change in the specific heat $\Delta C_v$ can now be readily calculated from $\Delta S$:
\begin{align}
\Delta C_v&=T\frac{\partial \Delta S}{\partial T}\bigg\vert_V\notag\\
&=-\frac{13\pi^4}{80} nk_B \left( B^\lambda
\log \vert\frac{T}{T_{\textrm{cut}}}\vert+\textrm{\foreignlanguage{russian}{B}}^\lambda 
\right) \left(\frac{T}{T_F}\right)^3-\frac{\log(2)\pi^2}{4}\frac{q_c}{p_F}nk_B \textrm{\foreignlanguage{russian}{B}}^\lambda  \left(\frac{T}{T_F}\right)^2 
\end{align}
Hence, simplifying and using our previous expressions for the factors $\textrm{\foreignlanguage{russian}{B}}^\lambda $ and $B^\lambda$, and limiting ourselves to a single scattering channel $\lambda$, we find the specific heat to be
\begin{align}
\frac{\Delta C_v}{nT} \approx \frac{\pi^4}{384} \log(2)\frac{k_B^2}{\epsilon_F} \frac{T_{\textrm{cut}}}{T_M}\frac{T}{T_F }(A_0^a)^3\left(
1+\frac{13\pi^2}{20}\frac{T}{T_{\textrm{cut}}}
\right)+\frac{39\pi^4}{160}\frac{k_B^2}{\epsilon_F}\frac{T^2}{T_F^2} \log\mid\frac{T}{T_{\textrm{cut}}}\mid (A_0^a)^2 
\end{align}
where we have defined the Majorana temperature as
\begin{align}
T_M=\frac{\alpha v_F}{2k_B}\rightarrow \alpha =\frac{2k_B T_M}{v_F}
\end{align}
in analogy to the Fermi temperature
\begin{align}
T_F=\frac{\epsilon_F}{k_B}=\frac{p_F^2}{2 k_B m}=\frac{p_F v_F}{2k_B}
\end{align}
and defined the cutoff temperature to be
\begin{align}
T_{\textrm{cut}}=\frac{\hbar v_F q_c}{2k_B}
\end{align}
This completes the zero-field expression for $\Delta C_v/nT$ given in the text.

\subsection*{\normalsize IV D. Non-analytic contributions to the specific heat in a Majorana liquid II: Finite external magnetic field}

Building off of our previous derivation of the specific heat, we will now consider the effects of turning on some finite external magnetic field.
Recall the change in the quasiparticle energy due to Majorana quasiparticle interactions:
\begin{align}
&\Delta \widetilde{\epsilon}\notag\\
&=\frac{1}{6\pi^2v_F^2} \bigg\{
-\frac{3}{2}(p-p_F)^2v_F^2 \left\{
\frac{3}{2} q_c\textrm{\foreignlanguage{russian}{b}}^\lambda - (p-p_F)\left(
 b^\lambda\log\mid
\frac{\max((p-p_F),\,T)}{q_c}
\mid+  
\textrm{\foreignlanguage{russian}{b}}^\lambda
\right)
\right\}\notag\\
&\phantom{=\frac{1}{6\pi^2 v_F^2}\bigg\{}\hspace{-12mm}+3\log(2) (k_B T)(p-p_F)v_F \left(q_c\textrm{\foreignlanguage{russian}{b}}^\lambda -(p-p_F)\left\{
 b^\lambda\log\mid
\frac{\max((p-p_F),\,T)}{q_c}
\mid+  
\textrm{\foreignlanguage{russian}{b}}^\lambda
\right\}
\right)\notag\\
&\phantom{=\frac{1}{6\pi^2 v_F^2}\bigg\{}
\hspace{-12mm}+\frac{\pi^2}{2}(k_B T)^2 (p-p_F)\left\{
 b^\lambda\log\mid
\frac{\max((p-p_F),\,T)}{q_c}
\mid+  
\textrm{\foreignlanguage{russian}{b}}^\lambda
\right\}\bigg\}
\end{align}
The effects of a magnetic field $M$ on the specific heat can be found by first calculating the dependence of $M$ on the density of states $\nu(T)$ and Landau parameter $f_0^a$, given as 
\begin{align}
M&=\nu(T)\left\{\left(\frac{1}{2}g \hbar\right)^2 H-f_0^a M\right\}+\frac{1}{2}\gamma \hbar \sum_{pp'}\Delta f\frac{\partial n}{\partial \epsilon}\delta n_p
\end{align}
where $g$ is the gyromagnetic ratio \cite{Carneiro1977Sep,Misawa1983Feb}. By using the fact that $M=\chi_H H$, the temperature-dependence of the susceptibility is contained within the change of the density of states at finite field. Finally, using the Maxwell relation
\begin{align}
H\frac{\partial^2 \chi}{\partial T^2}=\frac{\partial \gamma}{\partial H}
\end{align}
the $H$-field dependence of the specific heat $\gamma\equiv C(H,\,T)/T$ can be extracted.
Note that the integral over $dp$ in the calculation of the density of states will yield zero for any term proportional to $(p-p_F)^2$ or higher. Hence, we only care about the following terms: 
\begin{align}
&\frac{1}{6\pi^2v_F^2} \bigg\{3\log(2) (k_B T)(p-p_F)v_F q_c\textrm{\foreignlanguage{russian}{b}}^\lambda
\\
&\phantom{\frac{1}{6\pi^2v_F^2} \bigg\{}+\frac{\pi^2}{2}(k_B T)^2 (p-p_F)\left\{
 b^\lambda\log\mid
\frac{\max((p-p_F),\,T)}{q_c}
\mid+  
\textrm{\foreignlanguage{russian}{b}}^\lambda
\right\}\bigg\}
\end{align}
Simplifying, we have
\begin{align}
\Delta \epsilon_p\approx &\frac{\log(2)}{\pi^2 v_F^2} \frac{(k_B^2 T T_{\textrm{cut}})}{\hbar}(p-p_F)\textrm{\foreignlanguage{russian}{b}}^\lambda
\notag\\
&+\frac{1}{12v_F^2}(k_B T)^2 (p-p_F)\left\{
 b^\lambda\log\mid
\frac{\max((p-p_F),\,T)}{q_c}
\mid+  
\textrm{\foreignlanguage{russian}{b}}^\lambda
\right\}
\end{align}
The new, temperature-dependent density of states $\nu'(T)$ is therefore 
\begin{align}
\nu'(T)&=\nu(0)+\sum_p \frac{\partial n^0}{\partial \epsilon_p^0}\frac{\partial \Delta \epsilon_p}{\partial \epsilon_p^0}-\sum_{pp'}\Delta f_{pp'}\frac{\partial n^0}{\partial \epsilon_p^0}\frac{\partial n^0}{\partial \epsilon_{p'}^0}\notag\\
&\sim \nu(0)\left\{1+a T^2 b^\lambda \log\mid\frac{T}{T_{\textrm{cut}}}\mid+\beta \textrm{\foreignlanguage{russian}{b}}^\lambda T^2+\zeta \textrm{\foreignlanguage{russian}{b}}^\lambda T T_{\textrm{cut}}-G[b^\lambda,\,\textrm{\foreignlanguage{russian}{b}}^\lambda]\right\}
\end{align}
where $a\sim k_B^2/v_F^2$, $\beta$ and $\zeta$ are unitful constants, and $G$ is a function of $b^\lambda$ summed with $\textrm{\foreignlanguage{russian}{b}}^\lambda$. The susceptibility is then
\begin{align}
\chi&= \frac{\left(\frac{1}{2}g \hbar\right)^2 \nu'(T)}{1+f_0^\lambda \nu'(T)}\notag\\
&\approx \left(\frac{1}{2}g \hbar\right)^2 \nu(0)\bigg\{
\frac{(-1+G)}{-1+F_0^\lambda(-1+G)}+\frac{\textrm{\foreignlanguage{russian}{b}}^\lambda  T T_{\textrm{cut}} \zeta }{(-1+F_0^\lambda(-1+G))^2}
\notag\\
&\phantom{.............}+\frac{\textrm{\foreignlanguage{russian}{b}}^\lambda (-\beta-a\log(T/T_{\textrm{cut}})+F_0^s(-1+G)(\beta+a \log(T/T_{\textrm{cut}})) +\textrm{\foreignlanguage{russian}{b}}^\lambda T_{\textrm{cut}}^2 \zeta^2 )T^2}{
(-1+F_0^\lambda(-1+G))^3
}
\bigg\}
\end{align}
We will now take the second derivative of $\chi$ with respect to $T$:
\begin{align}
&\frac{\partial^2 \chi}{\partial T^2}
\approx \left(\frac{1}{2}g \hbar\right)^2\nu(0)\frac{\textrm{\foreignlanguage{russian}{b}}^\lambda  }{(-1+F_0^\lambda(-1+G))^3}\notag\\
&\hspace{-12mm}\phantom{\frac{\partial^2 \chi}{\partial T^2}
\approx} \times \left\{
-3a-2\beta+2\textrm{\foreignlanguage{russian}{b}}^\lambda T_{\textrm{cut}}^2 \gamma^2 -2a \log\left(\frac{T}{T_{\textrm{cut}}}\right)+(-1+G)\left(3a+2\beta+2a \log\left(\frac{T}{T_{\textrm{cut}}}\right)\right)F_0^\lambda
\right\}\notag
\end{align}
Taking $G\sim \delta(b^\lambda+\textrm{\foreignlanguage{russian}{b}}^\lambda)$ and assuming $\textrm{\foreignlanguage{russian}{b}}^\lambda$ is very large, the above becomes 
\begin{align}
\frac{\partial^2 \chi}{\partial T^2} &\approx \frac{ \left(\frac{1}{2}g \hbar\right)^2  }{(-1+F_0^\lambda (-1+\delta(B^\lambda+ \textrm{\foreignlanguage{russian}{B}}^\lambda))^3}
\bigg\{2\textrm{\foreignlanguage{russian}{b}}^\lambda T_{\textrm{cut}}^2 \gamma^2 +  F_0^\lambda\left(3a +2\beta+2a\textrm{\foreignlanguage{russian}{B}}^\lambda\log\left(\frac{T}{T_{\textrm{cut}}}\right)\right) \notag\\
&\phantom{\approx}\hspace{50mm}\times\left(-2+\delta(B^\lambda+ \textrm{\foreignlanguage{russian}{B}}^\lambda)\right)\bigg\}
\end{align}
For small values of $\textrm{\foreignlanguage{russian}{B}}^\lambda$, $\partial^2 \chi/\partial T^2$ becomes temperature independent. For larger values, we get
\begin{align}
\frac{\partial^2 \chi}{\partial T^2} 
&\sim  \left(\frac{1}{2}g \hbar\right)^2\frac{2a   }{(F_0^\lambda)^2 \delta^3 \textrm{\foreignlanguage{russian}{B}}^\lambda}\log\left(\frac{T}{T_{\textrm{cut}}}\right)  
\end{align}
Therefore, the specific heat should go as
\begin{align}
\gamma(H)=\gamma(0)+H^2 A \left(\frac{1}{2}g \hbar\right)^2\frac{1   }{(F_0^\lambda)^5}\frac{\alpha}{p_F}\log\left(\frac{T}{T_{\textrm{cut}}}\right)  
\end{align}
where we have used the fact that $ \textrm{\foreignlanguage{russian}{B}}^\lambda \sim \frac{\pi^2}{8}\frac{p_F}{\alpha} (F_0^\lambda)^3$.
As such, the weak coupling limit of the Majorana system (i.e., small $\alpha$ and $F_0^\lambda$) will result in an apparent lowering of the Sommerfeld coefficient, and thus we have derived the $T$-dependence of the specific heat at finite $H$ as discussed in the text.

\vspace{5mm}

\noindent {\it \bf Data availability}-- All data used in this work are deposited in a Google Drive folder and will be shared upon request by contacting the corresponding author.

\vspace{10mm}
\noindent {\it \bf Code availability}-- The python code used for the plots in this article are available upon reasonable request by contacting the corresponding author.

\vspace{5mm}
\noindent {\it \bf Acknowledgements}-- We thank Emilio Cobanera, Vincent Flynn, and Lorenza Viola for a thorough critique of this manuscript. Work at Boston College was supported by the John H. Rourke endowment fund (J.T.H. \& K.S.B.) and the U.S. Department of Energy, Office of Basic Energy Sciences, Division of Physical Behavior of Materials under Award No. DE-SC0023124 (F.B. \& F.T.). {Work in Los Alamos was supported by the U.S. Department of Energy, Office of Science, National Quantum Information Science Research Centers, and Quantum Science Center} (S.L. \& R.M.).
One of the authors (J.T.H.) would like to thank the organizers of the 2020 virtual Princeton Summer School on Condensed Matter Physics, where the initial results of this paper were presented.

\vspace{5mm}
\noindent {\it \bf Author contributions}-- J.T.H., X.C., and K.B. performed the theoretical calculations and contributed to the theoretical predictions and analysis of the experimental data. F.B. and F.T. prepared the samples and contributed to the analysis of the experimental data. S.L. and R.M. performed experimental measurements on the samples and contributed to the analysis of the experimental data. 

\vspace{5mm}
\noindent {\it \bf Competing interests}-- The authors declare no competing interests.


\bibliography{main.bib}{}

\end{document}